\documentclass[11pt, a4paper, twoside, openright]{memoir}
\usepackage[square,comma,numbers,sort&compress]{natbib}
\usepackage{graphicx,amsmath,bbm} 
\usepackage[breaklinks,bookmarks]{hyperref}
\usepackage{memhfixc}
\makepagestyle{custom}
\makeevenhead{custom}{\normalfont\thepage}{}{J.~Keeling,
M.~H.~Szyma\'nska, P.~B.~Littlewood}
\makeoddhead{custom}{\rightmark}{}{\normalfont\thepage}

\newcommand{\Tr}{{\mathrm{Tr}}}
\newcommand{\nodagger}{{\vphantom{\dagger}}}

\setstocksize{310mm}{210mm}
\settrimmedsize{310mm}{210mm}{*}
\settypeblocksize{*}{1.1\lxvchars}{1.8}
\setlrmargins{20mm}{*}{*}
\setulmargins{*}{*}{1}
\setmarginnotes{20pt}{110pt}{\onelineskip}
\checkandfixthelayout

\renewcommand{\thesection}{\arabic{section}}

\changecaptionwidth
\captiontitlefont{\small}
\captionwidth{0.8\textwidth}

\begin{document}

\bibliographystyle{apsrev}

\frontmatter
\pagenumbering{roman}

\pagestyle{empty}
\clearpage
\begin{center}

\large{{\textbf{Keldysh Green's function approach to coherence in a
non-equilibrium steady state: connecting Bose-Einstein condensation and lasing}}}

\vspace{10ex}

Jonathan Keeling,
%$^1$, 
Marzena H. Szyma\'nska and
%$^2$
Peter B. Littlewood
%$^1$ 
\\
\let\thefootnote\relax\footnotetext{ \hspace{-9mm} Jonathan Keeling \\ 
  Cavendish Laboratory, University of Cambridge, email: jmjk2@cam.ac.uk \\
  Marzena H. Szyma\'nska\\ Department of Physics, University of Warwick, email: M.H.Szymanska@warwick.ac.uk
  \\
  also at London Centre for Nanotechnology
  \\ Peter Littlewood\\ 
  Cavendish Laboratory, University of Cambridge, email: pbl21@cam.ac.uk }

\end{center}
\vspace{10ex}

\noindent\textbf{Abstract:}

\noindent
Solid state quantum condensates often differ from previous examples of
condensates (such as Helium, ultra-cold atomic gases, and
superconductors) in that the quasiparticles condensing have relatively
short lifetimes, and so as for lasers, external pumping is required to
maintain a steady state.  On the other hand, compared to lasers, the
quasiparticles are generally more strongly interacting, and therefore
better able to thermalise.  This leads to questions of how to describe
such non-equilibrium condensates, and their relation to equilibrium
condensates and lasers.  This chapter discusses in detail how the
non-equilibrium Green's function approach can be applied to the
description of such a non-equilibrium condensate, in particular, a
system of microcavity polaritons, driven out of equilibrium by
coupling to multiple baths.  By considering the steady states, and
fluctuations about them, it is possible to provide a description that
relates both to equilibrium condensation and to lasing, while at the
same time, making clear the differences from simple lasers.

\clearpage
\pagestyle{custom}
\mainmatter

Bose-Einstein condensation (BEC), the equilibrium phase transition of
weakly interacting bosons, was realised over ten years ago in
ultra-cold atomic gases. After long and strenuous efforts to observe
this state in solids, BEC of polaritons\cite{kasprzak06} and of
magnons\cite{demokritov06} were reported. These reports
followed observations of related effects for excitons in quantum Hall
bilayers\cite{eisenstein04}, spin triplet states in magnetic
insulators\cite{ruegg03} and excitons in coupled quantum
wells\cite{butov02A,butov02B,snoke02nature}.  Solid-state condensates
depart from the archetypal BEC in several ways. Most importantly they
live for short times and rely on external pumping.  Indeed, it was the
decay, and consequent lack of equilibrium, which for a long time
presented the obstacle the realisation of solid-state BEC.  Even if
one can accelerate thermalisation, the decay, and the consequent flux
of particles, remains a more important effect in solid state than it
generally does in cold atomic gases, or in other quantum condensates
such as superfluid Helium.

When considering whether such a system may be treated as equilibrium
or not, there are several distinct characterisations of the degree to
which the system is non-equilibrium.  The most obvious compares
particle lifetime to the time required for collisions to thermalise
the system, determining the extent to which a thermal distribution may
arise.  The timescale for establishing a thermal distribution within
one part of the system can however be quite different to that for
establishing either thermal or chemical equilibrium between different
parts of the system.  Another characterisation of whether
non-equilibrium physics is relevant arises from comparing the
linewidth due to finite particle lifetime to the temperature of the
system, thus determining whether lifetime or temperature effects
dominate coherence properties.  Table \ref{tab:energy-time-scales}
summarises the typical timescales and energy scales connected with
different examples of metastable quantum condensates.
It is clear that the ratio of thermalisation time to the particle
lifetime is generally somewhat larger for solid-state condensates than
it is for cold atomic gases.  If one instead compares the ratio of the
linewidth due to decay to the characteristic temperature, polaritons
stand out as having a decay linewidth of the same order of magnitude
as their temperature. As such, polaritons are good systems in which
to study effects of finite lifetime on coherence properties.
\begin{table*}[htpb]
  \centering
  \begin{tabular}{l|cc|ccc}
    &Lifetime
    &Thermalisation
    &Linewidth
    &\multicolumn{2}{c}{Temperature}
    \\
    \hline
    Atoms\cite{streed06} &
    10s &
    10ms &
    $2.5\times10^{-13}$meV &
    $10^{-8}$K &
    $10^{-9}$meV
    \\
    Excitons\cite{hammack07} &
    50ns &
    0.2ns & 
    $5\times 10^{-5}$meV &
    1K & 
    0.1meV
    \\
    Polaritons\cite{deng06:eqbm} &
    5ps&
    0.5ps & 
    0.5meV &
    20K& 
    2meV
    \\
    Magnons\cite{demokritov06} &
    1$\mu$s &
    100ns & 
    $2.5\times10^{-6}$meV &
    300K & 
    30meV
  \end{tabular}
  \caption[Characteristic timescales and energies]{Characteristic
    timescales and energies for: particle lifetimes, times to
    establish a thermal distribution, linewidth due to finite
    lifetime, and characteristic temperatures for various candidate
    condensates.  Comparison of the first two describes how thermal
    the distribution will be; comparison of the later two determine
    the effect of finite lifetime on coherence properties.}
  \label{tab:energy-time-scales}
\end{table*}

Because, as we will discuss further below, polariton condensates
provide such a clear illustration of the properties of non-equilibrium
condensation, we will focus on them in particular.
Microcavity polaritons are the quasiparticles which result from strong
coupling between photons confined in a semiconductor microcavity, and
excitons in a quantum well.
By changing the detuning between the excitons and photons, and by
changing the strength of an external pump that injects polaritons, one
can modify the polariton mass, density and the effect of interactions
between polaritons.
A more detailed introduction to microcavity polaritons and
semiconductor microcavities can be found in several review articles
and books
\cite{skolnick98,savona99,yamamoto00,ciuti03,special:05,keeling07:review,kavokin:oup}

The intrinsic non-equilibrium and dissipative nature of solid-state
condensates, especially of polaritons, brings connections to other
systems exhibiting macroscopic coherence, i.e lasers.  With the
realisation of more complex, interaction dominated lasers, such as
random lasers (see e.g. Refs.~\cite{cao05,tuereci08}) or atom lasers
(e.g. Refs.\cite{mewes97,bloch99,hagley99}), this connection is
particularly pronounced.
Compared to simple lasers, polaritons are however more strongly
interacting, and therefore much better able to thermalise than are
photons, and so in many ways solid-state condensates can be viewed as
being somewhere in between an equilibrium BEC and a laser.
At the same time, at large temperatures and/or in the presence of large
decoherence mechanisms and large pumping the same microcavity system
supports a simple lasing action.
In this context, microcavity polaritons provide particularly excellent
playground for studying coherence in a dissipative environment, and
the differences and similarities between condensates and lasers.
Clearly, an approach which takes into account the non-equilibrium and
dissipative nature of this new state of matter, as well as strong
interactions, multimode-structure, low dimensionality and finite size
is necessary.

This chapter will discuss a theoretical approach to modelling quantum
condensates that are driven out of equilibrium by a flow of particles
through the system.  
We therefore consider coupling the system to baths, which can transfer
energy as well as particles to and from the system.  With such baths,
we find that the behaviour of a simple laser can be recovered in the
limit of high temperature baths.  A different scenario of how
decoherence affects condensation can be found if one considers static
disorder --- i.e. allowing scattering, but with no transfer of energy
to or from the system.  Such a problem
\cite{szymanska02,szymanska03:pra} is closely related to the
Abrikosov-Gorkov approach to disordered
superconductors\cite{abrikosov_gorkov}.  As in the case of
superconductors, one finds a distinction between ``pair-breaking'' and
``non-pair-breaking'' disorder (respectively magnetic and non-magnetic
impurities in the superconducting case).  As expected from Anderson's
theorem\cite{anderson59}, the coherence associated with the condensate
leads to a gap in the exciton density of states, which makes the
condensate robust to non-pair-breaking disorder.  With pair-breaking
disorder, decoherence eventually destroys the gap and finally the
condensate, but for small amounts of decoherence, the gap protects the
condensate.  A similar scenario also exists in the ultra high density
limit, where excitons are destroyed by screening, leading to an
electron-hole plasma phase\cite{marchetti04:prb}, which can
nonetheless support lasing.  While we focus in this chapter instead on
the effects of particle flux, and baths that can transfer energy,
these other results illustrate that there are a variety of ways in
which decoherence can either suppress or modify the properties of a
condensate.  In principle one can have both a crossover from a
polariton condensate to a regular laser (weak coupling but still
excitonic gain medium, as discussed here), and a crossover to a
particle-hole laser (weak coupling, electron-hole plasma, if screening
is strong).

The approach to modelling the condensate with a flux of particles
presented in this chapter is based on work by the authors in
Refs.\cite{szymanska06:keldysh,szymanska07}.  While in those works,
the results were derived and presented making use of the
non-equilibrium path integral approach\cite{kamenev05}, both the
results and their theoretical basis can be understood without this
technical background, by considering the diagrammatic approach to
calculating non-equilibrium Green's functions
\cite{keldysh65,danielewicz84,lifshitz:Phys_Kin}.  The particular aim
of this chapter is therefore to review some of these results,
illustrating in some detail how a steady state non-equilibrium system
which develops spontaneous coherence, can be treated in the
non-equilibrium diagrammatic formalism.  At the same time, this
approach will provide a natural language to highlight the way this
system relates both to equilibrium condensates and to lasers, and to
understand the ingredients that makes it differ from these limits.

There are a number of other known approaches to describing systems
driven out of equilibrium by coupling to multiple baths. Those that
have been applied to microcavity polaritons include: quantum kinetic
equations\cite{tassone97:prb,tassone99,malpuech02:apl, malpuech02:prb,
  porras02,doan05:prb,doan06,doan08}, Heisenberg-Langevin
equations\cite{mieck02}, stochastic methods for density matrix
evolution (i.e. truncated Wigner
approxi\-mation)\cite{carusotto05,wouters09}; as well as mean-field
approaches, considering the complex Gross-Pitaevksii equation, in some
cases including also coupling to reservoirs or thermal baths
\cite{wouters07:bec,wouters08:bec,keeling08:gpe}.  While this chapter
does not intend to review the merits of each of these approaches, it
is worth noting that in general, these approaches are all connected.
The connections between many of them can simply be seen by looking at
their relation to the non-equilibrium diagrammatic approach.  As
discussed in \cite{lifshitz:Phys_Kin,kadanoff62}, the quantum
Boltzmann equation can be derived as an equation for the distribution
function that appears in the Keldysh Green's function, along with a
Wigner transformation from $F(t,r,t^\prime, r^\prime)$ to
$F(T,R,\omega, p)$.  It will become clear from the discussion in
Sec.~\ref{sec:decay-bath-langle}, that there is a close analogy
between the Keldysh Green's functions and the Heisenberg-Langevin
equations, with the bath Green's functions describing the same physics
as the correlation functions of the bath noise operators in the
Heisenberg-Langevin approach.
There also exists a connection between the approach described here and
density matrix evolution.  The single particle density matrix is given
by $\langle \psi^\dagger(r,t) \psi(r^\prime, t)\rangle$, and thus
corresponds to an appropriate combination of equal time Green's
functions.  The density matrix naturally gives single time expectations
of appropriate observables,  it is also possible to derive two-time
correlations from the time evolution of the density matrix, by means of
the quantum regression theorem\cite{scully97}.  The quantum regression
theorem however relies on making an additional Markov approximation
regarding the bath occupations, as well as a Markov approximation for
the bath density of states\cite{ford96}.  The Keldysh Green's function
approach does not require this additional Markov approximation;  and in
fact Sec.~\ref{sec:recov-laser-limit} will show how making this further
approximation restricts the conditions for condensation to occur.

In order to illustrate the application of the non-equilibrium
technique, we consider a specific model of microcavity polaritons,
starting from disorder localised excitons strongly coupled to cavity
photons\cite{eastham00:ssc,keeling04,marchetti06}.  In this model,
interactions between excitons are included by treating the excitons as
hard-core bosons, allowing one exciton, but no more, to occupy a given
disorder localised state.  For the discussion presented here, using
this model provides a number of technical advantages: it connects
closely to the idea of gain from two-level systems that is typically
used in models of simple lasers\cite{scully97}, making the comparison
to lasing straightforward; and it automatically includes
nonlinearity of the excitons, allowing this nonlinearity to be
described by the properties of the exciton representation, rather than
requiring higher order diagrammatic corrections.  In addition, in an
equilibrium situation, the mean-field theory of this model is known to
give a reasonable description of the critical temperature, except at
very low densities where fluctuation corrections become
important\cite{keeling04}.

This chapter is organised as follows;
section~\ref{sec:model-hamilt-coupl} introduces the model Hamiltonian,
and its coupling to baths.  Section~\ref{sec:modell-non-equil} then
describes the approach we will take to modelling this system,
reviewing some standard results of the non-equilibrium diagrammatic
technique that will be used later, and discussing the mean-field
approach we use to find the steady state.  In order to evaluate this
mean-field condition, it is necessary to determine the effects of the
baths on the system, by calculating particular self energy diagrams,
these self energies are presented in
section~\ref{sec:effects-baths-system}.
Section~\ref{sec:mean-field-theory-3} then discusses the mean-field
theory, considering how it can recover both equilibrium results
in one limit, as well as the description of a simple laser
in another limit.  Section~\ref{sec:fluct-inst-norm} discusses
fluctuations about the steady state, analysing stability, and further
illuminating the connection to (and distinctions from) a simple laser;
section~\ref{sec:fluct-cond-syst} then provides a more qualitative
discussion of the fluctuations of the condensed system, focusing in
particular on the combined effect of finite size and finite lifetimes.

\section{Polariton system Hamiltonian, and coupling to baths}
\label{sec:model-hamilt-coupl}

\begin{figure}[htpb]
  \centering
  \includegraphics[width=0.9\textwidth]{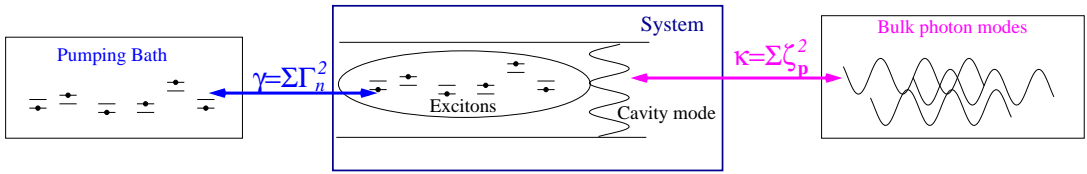}
  \caption{Cartoon of system, consisting of photons strongly coupled
    to excitons, and external baths, describing pumping and decay.
    Adapted from Ref.\cite{szymanska07}.}
  \label{fig:cartoon}
\end{figure}

As explained above, we consider a model of excitons as hard core
bosons coupled to propagating photons.  To write the Hamiltonian for
hard-core bosons, it is convenient to introduce fermionic operators
$b^\dagger_i, a^\dagger_i$, such that the two fermionic states
represent the presence or absence of an exciton on a given site, hence
the operator $b^\dagger_ia^\nodagger_i$ is the exciton creation
operator.  With this notation, the system is described by:
\begin{equation}
  \label{eq:1}
  H_{\mathrm{sys}} = \sum_i  \epsilon_i
    (b^\dagger_i b^\nodagger_i - a^\dagger_i a^\nodagger_i)
    + \sum_k \omega_k \psi^\dagger_k \psi^\nodagger_k
    + \sum_{i,k} g_i (\psi^\dagger_k a^\dagger_i b^\nodagger_i + \mathrm{H.c.}),
\end{equation}
where $\epsilon_i$ is the energy of a localised exciton state, $g_i$
is the exciton-photon coupling strength, and $\omega_k = \omega_0 +
k^2/2 m_{\mathrm{phot}}$ is the dispersion of cavity photons.  As
sketched in Fig.~\ref{fig:cartoon}, this can then be driven out of
equilibrium by coupling to two baths,
so the the system evolves under the full Hamiltonian 
$H = H_{\mathrm{sys}} + H_{\mathrm{sys,bath}} + H_{\mathrm{bath}}$.  It will be useful
later on to divide the coupling to baths into coupling to the pumping bath,
and coupling to the decay bath
$H_{\mathrm{sys,bath}} = H^{\mathrm{pump}}_{\mathrm{sys,bath}} + H^{\mathrm{decay}}_{\mathrm{sys,bath}}$
where the forms of the coupling to the pumping and decay baths are:
\begin{equation}
  \label{eq:2}
  H^{\mathrm{pump}}_{\mathrm{sys,bath}} = 
  \sum_{n,i} \Gamma_{n,i} 
  \left(a^\dagger_i A_n + b^\dagger_i B_n + \mathrm{H.c.} \right),
  \quad
  H^{\mathrm{decay}}_{\mathrm{sys,bath}} = 
  \sum_{p,k} \zeta_{p,k} \left(\psi^\dagger_k \Psi_p + \mathrm{H.c.}\right).
\end{equation}
Here $\Gamma_{n,i}$ is the coupling to a pumping bath, described by
the fermionic operators $B^\dagger_n, A^\dagger_n$, and $\zeta_{p,k}$
is the coupling to decay bath, describing bulk photon modes
$\Psi^\dagger_p$.   
The bath Hamiltonian is taken to have the simple quadratic form:
\begin{equation}
  \label{eq:2b}
  H_{\mathrm{bath}} = 
  \sum_n \nu_n^\Gamma \left(B^\dagger_n B^\nodagger_n - A^\dagger_n A^\nodagger_n\right)
  +
  \sum_p \omega_p^\zeta \Psi^\dagger_p \Psi_p
\end{equation}

Describing the pumping reservoir and photon decay as baths means that
we assume these both contain many modes (i.e. are much larger than the
system), and thermalise rapidly compared to the interaction with the
system.  These assumptions mean that one may impose a particular
distribution function on the bath modes, and then determine what
distribution the system adopts; we will take a thermal distribution
for the pumping bath, specified by a bath temperature and chemical
potential, and we will assume the bulk photon modes are unoccupied.
Note that we do not explicitly introduce any system chemical
potential, as the density of the system will be fixed by the balance
of pumping and decay, however a natural definition of the system
chemical potential will arise later.

\section{Modelling the non-equilibrium system}
\label{sec:modell-non-equil}

The Keldysh non-equilibrium diagrammatic technique\cite{keldysh65} is
an approach well suited to dealing with the kind of non-equilibrium
steady state which we consider here.
Section~\ref{sec:non-equil-diagr} briefly summarises the concepts that
will be important in the remainder of this chapter; for a more
complete introduction, see for example
Refs.~\cite{keldysh65,danielewicz84,lifshitz:Phys_Kin}.  Within this
diagrammatic approach, we will then determine the possible steady
states of the system by a mean-field approach, introduced in
section~\ref{sec:mean-field-condition}.

\subsection{Non-equilibrium diagram approach}
\label{sec:non-equil-diagr}

\begin{figure}[htpb]
  \centering
  \includegraphics[width=0.9\textwidth]{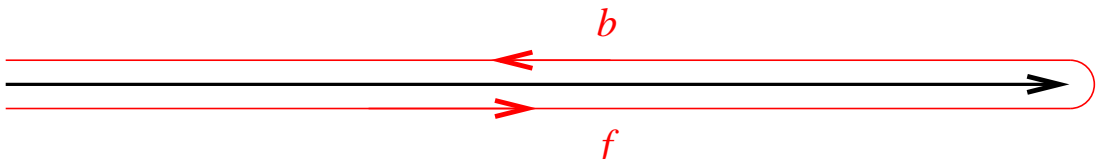}
  \caption{Keldysh closed-time-path contour, which can generate
    multiple orderings of fields}
  \label{fig:keldysh-contour}
\end{figure}

In order to determine both the spectrum (i.e. the ground and excited
states, taking into account interactions and coupling to baths), and
the non-equilibrium occupation of this spectrum, it is necessary to
calculate two linearly independent Green's functions; it is convenient
to make these the retarded and Keldysh Green's functions:
\begin{equation}
  \label{eq:3}
   D^R(t,r) = -i \theta(t) \left< [ \psi(t,r), \psi^\dagger(0,0)]_-\right>,
\ \
   D^K(t,r) = -i \left< [ \psi(t,r), \psi^\dagger(0,0)]_+ \right>.
\end{equation}
Here, $[\psi,\psi^\dagger]_{\mp}$ indicates the
commutator (anti-commutator) of $\psi$ and $\psi^\dagger$.
These Green's functions can be written as time-ordered products of
fields by introducing the Keldysh contour, shown in
Fig.~\ref{fig:keldysh-contour}.  
Each point on this contour is labelled by $(t,\{f,b\})$, where the
$f,b$ label whether it is on the forward or backward branch.  We then
introduce the contour time ordering $T_c$, such that fields on the
backward contour are always later than those on the forward contour,
and that pairs of fields on the backward contour should appear in
reverse order.  By then introducing symmetric and anti-symmetric
combinations of these fields $\psi_{\pm} = \left[\psi(t,f) \pm
  \psi(t,b)\right]/\sqrt{2}$, one may write the Green's function:
\begin{equation}
  \label{eq:4}
  D = 
  \left( 
    \begin{array}{cc}
        D^K & D^R \\ D^A & 0 
    \end{array}
  \right)
  =
  -i
  \left<
    T_c \left(
      \begin{array}{c}
        \psi_+(t,r) \\ \psi_-(t,r) 
      \end{array}
    \right)
    \left(
      \psi_+^\dagger(0,f),  \psi_-^\dagger(0,b) 
      \right)
  \right>.
\end{equation}
Here, $D^A$ refers to the advanced Green's function, which is the
Hermitian conjugate of the retarded Green's function.

Given the above time-ordered products, one may use standard
methods\cite{lifshitz,agd} to write a diagrammatic expansion, by
writing the Heisenberg picture fields in terms of the interaction
picture fields $\tilde{\psi}(t)$:
\begin{equation}
  \label{eq:5}
  \psi(t) = U^{-1}(t) \tilde{\psi}(t) U(t), \qquad
  \tilde{\psi}(t) = e^{i H_0 t} \psi e^{-i H_0 t},
\end{equation}
where $H=H_0 + H_{\mathrm{int}}$, and $H_0$ is ``free'', meaning
that it is simple to write expectations of products of fields
evolving according to $H_0$.  By formally solving the equation 
for $U(t)$, one may then write the Green's functions in the following
form:
\begin{eqnarray}
  \label{eq:6}
  D &=& 
  -i
  \left<
    T_c \left[
      \left(
        \begin{array}{c}
          \tilde{\psi}_+(t,r) \\ \tilde{\psi}_-(t,r) 
        \end{array}
      \right)
      \left(
        \tilde{\psi}_+^\dagger(0,0),  \tilde{\psi}_-^\dagger(0,0) 
      \right)
      U
    \right]
  \right> 
  \\
  \label{eq:7}
  U
  &=&
  \exp\left[-i \int_C  
    \tilde{H}_{\mathrm{int}}(t)
    dt \right]
  =
  \exp\left[-i \int_{-\infty}^\infty  \left(
      \tilde{H}_{\mathrm{int}}(t,f)
      -
      \tilde{H}_{\mathrm{int}}(t,b)
    \right)
    dt \right].
\end{eqnarray}
The diagrammatic expansion then follows by expanding the exponential,
which produces vertices coupling free fields, and connecting these
vertices by lines representing the Green's functions of the free
fields. Compared to other diagrammatic expansions, the only extra
complication is to keep track of the $\pm$ labels on the fields, both
in the matrix structure of Keldysh/retarded/advanced Green's functions,
and in the form of $U$.

In the following, we will frequently make use of the Dyson
equation\cite{agd,lifshitz,lifshitz:Phys_Kin},
${D}^{-1}_{\vphantom{0}} = {D}^{-1}_0 - \Sigma$, and so it is useful
to record the free inverse Green's function.  The inverse Green's
function has the structure:
\begin{equation}
  \label{eq:8}
 D^{-1} = \left[
  \left(
    \begin{array}{cc}
      {D}^{K} & {D}^{R} \\
      {D}^{A} & 0
    \end{array}
  \right)\right]^{-1}
  =
  \left(
    \begin{array}{cc}
      0 & \left[{D}^{A}\right]^{-1} \\
      \left[{D}^{R}\right]^{-1} & \left[{D}^{-1}\right]^{K}
    \end{array}
  \right),
\end{equation}
where $\left[{D}^{-1}\right]^{K} = - \left[{D}^{R}\right]^{-1} {D}^K
\left[{D}^{A}\right]^{-1}$.  Using the results for a free field, one has
\begin{equation}
  \label{eq:9}
  \left[{D}_0^{R}\right]^{-1} = \omega - \omega_k + i \eta, \qquad
  \left[{D}_0^{-1}\right]^{K} = (2 i \eta) (2 n_B(\omega) + 1),
\end{equation}
where $\eta$ is infinitesimal.   All of the results noted above assume
bosonic fields;  the results for fermionic fields are similar, but commutators
and anti-commutators are interchanged in the definitions of Keldysh and
retarded Green's functions.

For our particular model of microcavity polaritons, the division of the
full Hamiltonian into $H_0$ and $H_{int}$ will be to take:
\begin{equation}
  \label{eq:9b}
    H_0 = 
  \sum_i  \epsilon_i (b^\dagger_i b^\nodagger_i - a^\dagger_i a^\nodagger_i)
  +
  \sum_k  \omega_k \psi^\dagger_k \psi^\nodagger_k
  +
  \sum_{i} g_i \psi_0
  \left(
    e^{i \mu_S t} a^\dagger_i b^\nodagger_i + 
    e^{-i \mu_S t}  b^\dagger_i  a^\nodagger_i
  \right) + H_{\mathrm{bath}}
\end{equation}
where $\psi_0$ is a mean-field coherent photon field, as discussed in
the next section.  This means that $H_{int}$ will contain the
system--bath interactions, as well as the interaction between the
two-level systems and incoherent photon fluctuations.  In the
following we will however generally focus on one part of $H_{int}$ at
a time.

\subsection{Mean-field condition for coherent state}
\label{sec:mean-field-condition}

For a system coupled to multiple baths, the mean-field theory can no
longer be thought of as minimising free energy, but rather as a stable
self consistent steady state. For a condensed solution, one looks for
a steady state of the form $\langle \psi_k \rangle = \psi_0
\exp(-i \mu_S t) \delta_{k,0} = \psi_0(t) \delta_{k,0}$, where $\mu_S$ is
introduced here merely as part of the steady state ans\"atz, but it will
be seen to play a role analogous to the equilibrium chemical
potential.  To be a self-consistent solution, this ans\"atz must satisfy the
Heisenberg equation: $\langle i \partial_t \psi \rangle = \langle [
\psi, H ] \rangle$, and so:
\begin{equation}
  \label{eq:10}
  \mu_S \psi_0(t) = \omega_0 \psi_0(t)
  + \sum_i g_i \langle a^\dagger_i(t) b^\nodagger_i(t) \rangle
  + \sum_p \zeta_{p,0} \langle \Psi_p(t) \rangle.
\end{equation}
The expression $\langle a^\dagger_i(t) b^\nodagger_i(t) \rangle$ describes
the polarisation of the two-level systems, and can be written
in terms of the Keldysh Green's function, as:
\begin{equation}
  \label{eq:11}
  \left< a^\dagger_i(t) b^\nodagger_i(t) \right> 
  =
  \frac{1}{2} \left< \left[ b^\nodagger_i(t), a^\dagger_i(t) \right]_- \right>
  =
  \frac{i}{2} G^K_{a^\dagger_i b^\nodagger_i}(t,t) = 
  \frac{i}{2} \int \frac{d \nu}{2 \pi} G^K_{a^\dagger_i b^\nodagger_i}(\nu).
\end{equation}
As well as this self-consistency condition to determine the coherent
field amplitude and the effective system chemical potential $\mu_S$,
the mean-field approach can also be used to give an estimate of the
polariton density.  This density will be used later in producing the
phase diagram of the polariton condensate.  The mean-field estimate of
the total density is given by the combination of the photon density
$|\psi_0|^2$, and the fermion density $(i/2)\Tr[G^K_{b^\dagger_i
  b^\nodagger_i} - G^K_{a^\dagger_i a^\nodagger_i}]$.

\section{Effects of baths on system correlation functions}
\label{sec:effects-baths-system}

In the above, we found that the mean-field condition could be written
in terms of the two-level system Green's function, and the expectation
of the decay bath fields.  In this section we will discuss in detail
the treatment of the baths and their effect on system's correlation
functions, which will then determine the conditions under which a
condensed solution may exist.  Most of the effort, in
Sec.~\ref{sec:pump-bath-gk_bd}, will be dedicated to finding
$G^K_{a^\dagger_i b^\nodagger_i}(\nu)$ including the effects of
pumping.  Before doing this, section~\ref{sec:decay-bath-langle} will
address the simpler problem of how $\langle \Psi_p(t)\rangle$ can be
related to the decay bath Green's function and thus evaluated.

\subsection{Decay bath and $\langle \Psi_p \rangle$}
\label{sec:decay-bath-langle}

To calculate $\langle \Psi_p(t)\rangle$ in terms of non-equilibrium Green's
functions, one may first use the interaction picture, in terms
of the system-bath coupling, to write
\begin{math}
  \Psi_p(t) = U^{-1}(t) \tilde{\Psi}_p(t) U(t)
\end{math}. Here, $U(t)$ is the time-ordered exponential as in
Eq.~(\ref{eq:5}):
\begin{equation}
  \label{eq:14}
  U(t) = {T} \exp\left[ -
    i \int_{-\infty}^t dt^\prime \tilde{H}^{\mathrm{decay}}_{\mathrm{sys,bath}}(t^\prime)
  \right].
\end{equation}
Then, consider  inserting a factor:
\begin{equation}
  \label{eq:27}
  1 = 
  {T} \exp\left[ 
    i \int_{t}^\infty dt^\prime \tilde{H}^{{\mathrm{decay}}}_{\mathrm{sys,bath}}(t^\prime)
  \right] \cdot
  {T} \exp\left[ -
    i \int_{t}^\infty dt^\prime \tilde{H}^{{\mathrm{decay}}}_{\mathrm{sys,bath}}(t^\prime)
  \right],
\end{equation}
either before or after $\tilde{\Psi}_p(t)$.  The resulting expression
implies that one has:
\begin{equation}
  \label{eq:73}
  \langle \Psi_p(t)\rangle
  =
  \langle T_C [\tilde{\Psi}_p(t,f) U] \rangle
  =
  \langle T_C [\tilde{\Psi}_p(t,b) U] \rangle
  =
  \frac{1}{\sqrt{2}}
  \langle T_C [\tilde{\Psi}_{p,+}(t) U] \rangle,
\end{equation}
where the last equality has made use of the fact that if the expectation
of $\tilde{\Psi}(t,f)$ and $\tilde{\Psi}(t,b)$ match, then the expectation
of $\tilde{\Psi}_-(t)$ must vanish.
We are interested in particular in the value of this expectation
$\langle \Psi_p \rangle$
when we consider the system in the mean-field approximation. In this case
the system bath interaction term is given by:
\begin{equation}
  \int_C dt \tilde{H}^{{\mathrm{decay}}}_{\mathrm{sys,bath}}(t)
  = 
  \int_{-\infty}^{\infty} \!\!\!\!dt
  \sum_p \zeta_{p,0} 
  \sqrt{2}
  [\tilde{\Psi}_{p,-}^\dagger(t) \psi_0(t)  +
  \psi^\ast_0(t) \tilde{\Psi}_{p,-}^\nodagger(t)].
\end{equation}
With vertices given by this interaction, the set of diagrams involved
in evaluating Eq.~(\ref{eq:73}) is particularly simple: the only
possible connected diagram is one with a single bath Green's function
connecting the source term in $H_{\mathrm{sys,bath}}$ to the field
$\tilde{\Psi}_{+}$ that we want to measure.  As such, the sum appearing in
Eq.~(\ref{eq:10}) can be written as:
\begin{equation}
  \label{eq:13}
  \sum_p \zeta_{p,0} \langle \Psi_p(t) \rangle
  =
  \sum_p \zeta_{p,0}^2 \int dt^\prime 
  {D}^R_{\Psi_p^\dagger\Psi^\nodagger_p}(t,t^\prime) \psi_0(t^\prime).
\end{equation}
The simple form this equation takes is also the form one would find by
making the Born approximation; i.e. assuming that $\zeta_{p,0}$ is
small, so that terms like $\sum_p \zeta_{p,0}^2$ should be kept, and
neglecting terms involving any higher power of $\zeta_{p,0}$.
However, in the current case, because of the linearity of the
coupling, no other connected diagrams exist, so no assumption of
smallness is required in order to neglect higher order terms.

For a free bath, one may write $\tilde{\Psi}_p(t) =
e^{-i\omega^\zeta_p t} \Psi_p$, and so the bath Green's function  is given
by
\begin{math}
  {D}^R_{\Psi_p^\dagger\Psi^\nodagger_p}(t,t^\prime) 
  = -i \theta(t-t^\prime) e^{-i \omega^\zeta_p(t-t^\prime)}.
\end{math}
Taking a Markovian approximation for the bath density of states and
coupling [i.e assuming that the product of the bath density of states
$N^\zeta(\omega)$ and the square of the system--bath coupling
$\zeta_{p,0}$ are constant: $\pi\zeta_{p,0}^2
N^{\zeta}(\omega)=\kappa$] then gives:
\begin{equation}
  \label{eq:15}
  \sum_p \zeta_{p,0}^2 e^{-i \omega^\zeta_p(t-t^\prime)} = 2 \kappa \delta(t-t^\prime).
\end{equation}
Putting this all together, the net effect of the decay bath on the
self-consistency equation is just to add a decay term, so the result
may be written as:
\begin{equation}
  \label{eq:16}
  (\omega_0 - \mu_S - i \kappa) \psi_0 e^{-i \mu_S t} =
  - \sum_i 
  \frac{i g_i }{2}
  \int \frac{d \nu}{2 \pi} G^K_{a^\dagger_i b^\nodagger_i}(\nu).
\end{equation}

\subsection{Pumping bath and ${G}^K_{a^\dagger b}$}
\label{sec:pump-bath-gk_bd}

The remaining task is to find the matrix of fermionic Green's
functions in the four by four space resulting from the $a,b$ fermionic
fields and the $\pm$ space associated with the closed-time-path
contour.  As above, we will take the interaction Hamiltonian to be the
coupling between the system and the bath.  This leaves the free
fermion Hamiltonian:
\begin{equation}
  \label{eq:17}
  H^{{\mathrm{TLS}}}_0 = 
  \sum_i  \epsilon_i (b^\dagger_i b^\nodagger_i - a^\dagger_i a^\nodagger_i)
  + \sum_{i} g_i \psi_0
  \left(
    e^{i \mu_S t} a^\dagger_i b^\nodagger_i + 
    e^{-i \mu_S t}  b^\dagger_i  a^\nodagger_i
  \right).
\end{equation}
It is possible to diagonalise this Hamiltonian by a unitary transformation,
and thereby write the appropriate free Green's function, however it is
first necessary to remove the time dependence introduced by the form
of the ans\"atz for the coherent field.  This can be achieved by
a gauge transformation:
\begin{equation}
  \label{eq:18}
  H \to H - \frac{\mu_S}{2} 
  \left[
    \sum_i \left(b^\dagger_i b^\nodagger_i- a^\dagger_i a^\nodagger_i\right)
    +
    \sum_n \left(B^\dagger_n B^\nodagger_n- A^\dagger_n A^\nodagger_n\right)
  \right].
\end{equation}
such that $b \to b e^{-i \mu_S t/2}, a \to a e^{i \mu_S t/2}$, which
removes the time dependence of the mean-field photon to fermion
coupling.  The gauge transformation for the bath modes that also
appears in Eq.~(\ref{eq:18}) is necessary to ensure no time dependence
is introduced into the system-bath coupling terms.  The net result is
to replace $\epsilon_i \to \tilde{\epsilon}_i = \epsilon_i - \mu_S/2$
in $H^{{\mathrm{TLS}}}_0$, and to shift the bath Green's function in frequency by $\pm
\mu_S/2$.

After the above transformation, the Hamiltonian can be diagonalised by
the unitary transformation:
\begin{equation}
  \label{eq:19}
    \left(
    \begin{array}{c}
      b_i \\ a_i
    \end{array}    
  \right)
  =
  \left(
    \begin{array}{rr}
      \cos(\theta_i) & \sin(\theta_i) \\
      - \sin(\theta_i) & \cos(\theta_i)
    \end{array}
  \right)
  \left(
    \begin{array}{c}
      \beta_i \\ \alpha_i
    \end{array}    
  \right),
\end{equation}
after which the free Hamiltonian takes the form $H^{{\mathrm{TLS}}}_0 = \sum_i E_i
(\beta^\dagger_i \beta^\nodagger_i - \alpha^\dagger_i
\alpha^\nodagger_i)$, where $\tan(2 \theta_i) = - g_i \psi /
\tilde{\epsilon}_i$ and $E_i^2 = \tilde{\epsilon}_i^2 + g_i^2 \psi_0^2$.
Since the Hamiltonian is diagonal in the $\beta,\alpha$ basis, the
retarded Green's functions in that basis are just $[\nu\mp E_i +
i\eta]^{-1}$ (where $\eta$ is infinitesimal), and so the retarded
Green's functions in the $b,a$ basis can be written as:
\begin{eqnarray}
  \label{eq:20}
  {G}^R_0(\nu)
  &=&
  \left(
    \begin{array}{rr}
      \cos\theta_i & \sin\theta_i \\
      - \sin\theta_i & \cos\theta_i
    \end{array}
  \right)
  \left(
    \begin{array}{cc}
      [\nu - E_i + i\eta]^{-1} & 0 \\ 0 & [\nu + E_i + i\eta]^{-1}
    \end{array}
  \right)
  \left(
    \begin{array}{rr}
      \cos\theta_i & -\sin\theta_i \\
      \sin\theta_i & \cos\theta_i
    \end{array}
  \right)
  \nonumber\\
  &=&
  \frac{1}{(\nu+i\eta)^2 - E_i^2} 
  \left(
    \begin{array}{cc}
      \nu + \tilde{\epsilon_i} + i \eta & g_i \psi_0 \\
      g_i\psi_0 & \nu - \tilde{\epsilon_i} + i \eta
    \end{array}
  \right)
  \\
  \left[{G}^R_0\right]^{-1}
  &=&
  \left(
    \begin{array}{cc}
      \nu - \tilde{\epsilon_i} + i \eta& -g_i \psi_0 \\
      -g_i\psi_0 & \nu + \tilde{\epsilon_i} + i \eta
    \end{array}
  \right).
\end{eqnarray}

Just as for the free bosonic Green's functions described in
Eq.~(\ref{eq:9}), $[G^{-1}_0]^K$ is infinitesimal.  Since coupling to
the pumping baths, which is discussed next, will add a
non-infinitesimal Keldysh self energy, we will neglect this
infinitesimal contribution.  Therefore, the expression for
$\left[{G}^R_0\right]^{-1}$ and its Hermitian conjugate are all that
is needed of the free Green's function. What remains is to determine
the self energy that arises from coupling to the pumping bath

Taking the part of the interaction Hamiltonian due to
$H^{{\mathrm{pump}}}_{\mathrm{sys,bath}}$, and inserting it
into the definition of $U$ in Eq.~(\ref{eq:7}), one has that the
interaction vertices are generated by:
\begin{eqnarray}
  \label{eq:21}
  \int_C dt \tilde{H}^{{\mathrm{pump}}}_{{\mathrm{sys,bath}}} 
  &=&
  \int_{-\infty}^\infty dt
  \left[ 
    \tilde{a}^\dagger_{i}(t,f) \tilde{A}^\nodagger_{n}(t,f)
    - 
    \tilde{a}^\dagger_{i}(t,b) \tilde{A}^\nodagger_{n}(t,b)
    + \ldots
  \right]
  \nonumber\\
  &=&
  \int_{-\infty}^\infty dt
  \left[ 
    \tilde{a}^\dagger_{i+}(t) \tilde{A}^\nodagger_{n-}(t)
    +
    \tilde{a}^\dagger_{i-}(t) \tilde{A}^\nodagger_{n+}(t)
    + \ldots
  \right].
\end{eqnarray}
As each pumping bath couples either to only $a$ modes or to only $b$
modes, there are no off diagonal self energy terms in the $a,b$ basis.
One may therefore concentrate first on $\Sigma_{a^\dagger a}$, and
$\Sigma_{b^\dagger b}$ will follow by analogy.  Written as a matrix in
the Keldysh space as defined in Eq.~(\ref{eq:6}), one has:
\begin{equation}
  \label{eq:22}
  \Sigma_{a^\dagger a}
  =
  \left(
    \begin{array}{cc}
      \Sigma_{a^\dagger a}^{++} & \Sigma_{a^\dagger a}^{+-} \\
      \Sigma_{a^\dagger a}^{-+} & \Sigma_{a^\dagger a}^{--}
    \end{array}
  \right).
\end{equation}
The $\pm$ labels determine the label of the incoming/outgoing fields,
and it is clear from Eq.~(\ref{eq:21}) that $+$ fields couple to $-$
bath fields and vice versa.  Thus, an example self energy diagram is:
\raisebox{-12pt}{\includegraphics{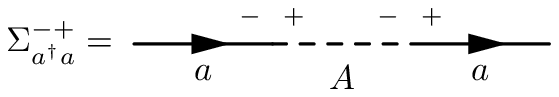}},
giving the equations
\begin{eqnarray*}
  \Sigma_{a^\dagger a}^{++}(t,t^\prime) &=&
  \sum_n \Gamma_{i,n}^2 G^{--}_{A^\dagger A} 
  = 0
  \\
  \Sigma_{a^\dagger a}^{-+}(t,t^\prime) &=& 
  \sum_n \Gamma_{i,n}^2 G^{+-}_{A^\dagger A} 
  = 
  - i\sum_n \Gamma_{i,n}^2 \theta(t-t^\prime) e^{-i\nu^\Gamma_n(t-t^\prime)}
  =-i \gamma \delta(t-t^\prime)
  \\
  \Sigma_{a^\dagger a}^{+-}(t,t^\prime) &=& 
  \sum_n \Gamma_{i,n}^2 G^{-+}_{A^\dagger A} 
  = 
  + i\sum_n \Gamma_{i,n}^2 \theta(t^\prime-t) e^{+i\nu^\Gamma_n(t-t^\prime)}
  =+i \gamma \delta(t-t^\prime)
  \\
  \Sigma_{a^\dagger a}^{--}(t,t^\prime) &=& 
  \sum_n \Gamma_{i,n}^2 G^{++}_{A^\dagger A} 
  = 
  - i\sum_n \Gamma_{i,n}^2 [1- 2n_A(\nu^\Gamma_n)] 
  e^{-i\nu^\Gamma_n(t-t^\prime)}
  =-2i \gamma \breve{F}_A(t-t^\prime).
\end{eqnarray*}
In the last three lines, the Markovian limit has been taken to give
the final equality.  In the last line we have used:
\begin{equation}
  \label{eq:23}
  \breve{F}_A(t) = 
  \int \frac{d\nu}{2\pi} e^{-i \nu t} F_A(\nu),
  \qquad
  F_A(\nu) = 1 - 2 n_A(\nu),
\end{equation}
where the form of the distribution function $F$ comes from the
form of the equal-time Keldysh Green's function $F_A(\nu_n^\Gamma) = 
\langle A^\nodagger_n A^\dagger_n - A^\dagger_n A^\nodagger_n \rangle$.
As a function of frequency, the self energy matrix in Keldysh space is thus:
\begin{equation}
  \label{eq:24}
  \Sigma_{a^\dagger a}(\nu)
  =
    \left(
    \begin{array}{cc}
      0 &  i \gamma \\ 
      -i \gamma & - 2 i \gamma F_A(\nu)
    \end{array}
  \right).
\end{equation}
The matrix for $\Sigma_{b^\dagger b}(\nu)$ is identical except that
$F_A(\nu) \to F_B(\nu)$.

Combining  the free Green's function and self energy, we may write the
entire inverse Green's function in the basis $(b_+, a_+, b_-, a_-)$ as:
\begin{equation}
  \label{eq:25}
  G^{-1}(\nu)
  =
  \left(
    \begin{array}{cccc}
      0 & 0 & \nu - \tilde{\epsilon_i} - i \gamma & - g_i \psi_0 \\
      0 & 0 & - g_i \psi_0 & \nu + \tilde{\epsilon_i} - i \gamma \\
      \nu - \tilde{\epsilon_i} + i \gamma & - g_i \psi_0 & 2 i \gamma F_B(\nu) & 0 
      \\
      - g_i \psi_0 & \nu + \tilde{\epsilon_i} + i \gamma &  0 & 2 i \gamma F_A(\nu)
    \end{array}
  \right).
\end{equation}
Clearly, if $\psi_0$ is zero then the $a,b$ fields decouple as
expected.  However for non-zero $\psi_0$, the Keldysh Green's
functions of the two fields get mixed, so that their occupation is
set by a balance of the pumping and the effects of the coherent photon
field.

To complete our analysis, we should invert the above matrix to find
the Keldysh block, and the $a^\dagger b$ component of that
block. Using the Keldysh block structure of Eq.~(\ref{eq:8}), and in
particular that $G^K=-G^R[G^{-1}]^K G^A$, one may write:
\begin{eqnarray}
  \label{eq:26}
  \lefteqn{{G}^K(\nu) =
  - \frac{ 2 i \gamma}{[(\nu+i\gamma)^2-E_i^2][(\nu-i\gamma)^2-E_i^2]}}
  \nonumber\\&&\times
  \left(
    \begin{array}{cc}
      \nu - \tilde{\epsilon_i} + i \gamma & - g_i \psi_0 \\
      - g_i \psi_0 & \nu + \tilde{\epsilon_i} + i \gamma \\
    \end{array}
  \right)
  \left(
    \begin{array}{cc}
      F_B(\nu) & 0 \\ 0 & F_A(\nu)
    \end{array}
  \right)
  \left(
    \begin{array}{cc}
      \nu - \tilde{\epsilon_i} - i \gamma & - g_i \psi_0 \\
      - g_i \psi_0 & \nu + \tilde{\epsilon_i} - i \gamma \\
    \end{array}
  \right).
\end{eqnarray}
For the mean-field condition in Eq.~(\ref{eq:10}), we require
in particular the  $G^K_{a_i^\dagger b^\nodagger_i}$ component which has the form:
\begin{equation}
  G^{K}_{a^\dagger_i b^\nodagger_i}(\nu)= 2i\gamma g\psi_0 
  \frac{[F_A(\nu)+F_B(\nu)]\nu
    +[F_B(\nu)-F_A(\nu)](\tilde{\epsilon}_i+i\gamma)}
  {[(\nu-E_i)^2+\gamma^2][(\nu+E_i)^2+\gamma^2]}.
  \label{eq:Gab}
\end{equation}

\section{Mean-field theory and its limits}
\label{sec:mean-field-theory-3}

Putting the $G^K_{a^\dagger b}$ component of Eq.~(\ref{eq:Gab}) into
Eq.~(\ref{eq:16}) gives the self-consistency condition (equation for
the condensate) of the mean-field theory:
\begin{equation}
  \label{eq:main}
  (\omega_0 - \mu_S - i \kappa) \psi_0  =
  \sum_i g^2_i \psi_0 \gamma \int \frac{d\nu}{2\pi}
  \frac{(F_B+F_A)\nu + (F_B-F_A)(\tilde{\epsilon}_i+i\gamma)}{%
     [(\nu-E_i)^2+\gamma^2][(\nu+E_i)^2 + \gamma^2]}.
\end{equation}
Equation (\ref{eq:main}) is central to our analysis.  This equation is
rather powerful, in that it combines several well known theoretical
results within a single framework.  As will be shown in section
\ref{sec:mean-field-theory-2}, in the equilibrium limit (where the
system--bath couplings are taken to zero) Eq.~(\ref{eq:main}) reduces
to the gap equation which applies throughout the BCS--BEC crossover.
In the opposite highly non-equilibrium limit (see section
\ref{sec:mean-field-theory}) it reduces to the standard laser
condition.  At low densities it reduces to the (complex)
Gross-Pitaevskii equation, discussed in section
\ref{sec:local-dens-appr}.  As such, this approach highlights the
connections between these apparently different descriptions of
condensates or lasers.

\begin{figure}[htpb]
  \centering
  \includegraphics[width=0.8\textwidth]{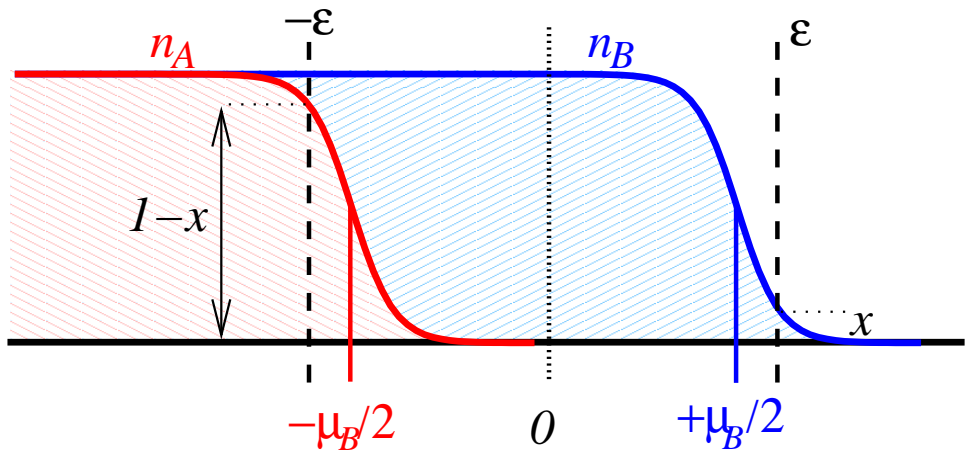}
  \caption[Occupation of pumping baths]{Occupation functions for the
    pumping baths, chosen to set total occupation of two modes to
    one, while varying the degree of inversion.}
  \label{fig:fermion-baths}
\end{figure}

The functions $F_{A,B}$ appearing in Eq.~(\ref{eq:main}) were defined
as $F_{A,B} = 1 - 2n_{A,B}$, where $n_{A,B}$ are bath occupation
functions.  These occupations are taken to be externally imposed, and
can be chosen to have any form relevant to a particular physical
situation. Here, we will choose these to be thermal and at equal
temperatures but different chemical potentials.  Noting that the
fermionic states were supposed to represent two-level systems (or
excitons), we take the occupations to satisfy $n_A+n_B=1$.  This
therefore requires that we have:
\begin{equation}
  \label{eq:28}
  F_{A,B}(\nu)=\tanh\left[
    \frac{\beta}{2} \left(
      \nu \pm \frac{\mu_B-\mu_S}{2} 
    \right)
  \right],
\end{equation}
where $\mu_B$ is an adjustable pumping bath chemical potential, and
$\mu_S$ appears in this expression due to the shift arising from the
gauge transformation in Eq.~(\ref{eq:18}). Schematically, this
situation is illustrated in Fig.~\ref{fig:fermion-baths}; one can see
that $F_A(-\epsilon) + F_B(\epsilon) = 2[1-n_A(-\epsilon) -
n_B(\epsilon)]=0$. Physically, this pumping process is most closely
related to electrical pumping.  Note that, in the absence of any other
processes, contact between the two-level systems (excitons) and the
pumping reservoir would control the population of the two-level
systems, and so one would have:
\begin{displaymath}
  \langle b^{\dagger}b-a^{\dagger}a \rangle
  =
  n_B(\epsilon)-n_A(-\epsilon)
  =
  -\mathrm{tanh}
  \left[ \frac{\beta}{2}\left(
        \epsilon-\frac{\mu_B}{2}+\frac{\mu_S}{2}
      \right)\right].
\end{displaymath}
Thus, by pumping with a thermalised bath, one will find a thermalised
distribution of excitons. Therefore, in the context of polaritons this
pumping scheme resembles closely pumping from a thermalised excitonic
reservoir, which is often the case in the experiments.

\subsection{Equilibrium limit of Mean-field theory}
\label{sec:mean-field-theory-2}

The simplest limit to recover from the non-equilibrium
self-consistency equation is that of thermal equilibrium.  This
corresponds to taking $\gamma, \kappa \to 0$.  Since the
self-consistency equation included only the coupling between
mean-field photons and the decay bath, there is no way that a thermal
distribution can be set by the decay bath.  On the other hand, the
pumping bath can set a thermal distribution, so to recover a
non-trivial equilibrium distribution one should take $\kappa \to 0$
first, and then $\gamma \to 0$.  If $\kappa = 0$, then the imaginary
part of the right hand side of Eq.~(\ref{eq:main}) must vanish.  In
order to satisfy this, without restricting the range of solutions of
the real part, one must choose $F_B(\nu) = F_A(\nu)$.  In terms of the
distribution functions written in Eq.~(\ref{eq:28}), this clearly
means $\mu_S=\mu_B$.  Physically, this means that in the absence of
decay, the chemical potential of the condensate matches the pumping
bath.

After fixing $\mu_S$, the remaining part of the equation becomes:
\begin{equation}
  \label{eq:29}
  (\omega_0 - \mu_B) \psi_0  =
   \sum_i g^2_i \psi_0 \gamma \int \frac{d\nu}{2\pi}
     \frac{2\tanh\left({\beta\nu}/{2} \right)\nu}{%
     [(\nu-E_i)^2+\gamma^2][(\nu+E_i)^2 + \gamma^2]}.
\end{equation}
We may then take the limit of small $\gamma$, by using:
\begin{equation}
  \label{eq:30}
  \lim_{\gamma \to 0}
  \frac{2 \gamma \nu}{[(\nu-E_i)^2+\gamma^2][(\nu+E_i)^2 + \gamma^2]}
  =
  \frac{2\pi}{4E_i} \left[ \delta(\nu-E_i) - \delta(\nu+E_i)\right],
\end{equation}
hence we find:
\begin{eqnarray}
  \label{eq:31}
  (\omega_0 - \mu_B) \psi_0  &=&
  \sum_i \frac{g^2_i \psi_0}{4E_i}  \int {d\nu}
  \tanh\left(\frac{\beta\nu}{2} \right)
  \left[ \delta(\nu-E_i) - \delta(\nu+E_i)\right]
  \nonumber\\
  &=&
  \sum_i \frac{g^2_i \psi_0}{2E_i}   \tanh\left(\frac{\beta E_i}{2} \right).
\end{eqnarray}
This is the equilibrium
result\cite{eastham00:ssc,keeling04,marchetti06}, but with the
two-level constraint on the fermions imposed only on
average.
\footnote{Imposing the two-level constraint on average, the
  equilibrium expectation of the inversion $\langle b^\dagger b -
  a^\dagger a \rangle$ can be written as:
\begin{equation}
\label{eq:32}
  \frac{ e^{\beta E} - e^{-\beta E}}{%
    1 + e^{\beta E} + e^{-\beta E} + 1} = 
  \frac{(e^{\beta E/2} - e^{-\beta E/2})(e^{\beta E/2} + e^{-\beta E/2})}{%
    (e^{\beta E/2} + e^{-\beta E/2})^2}
  = \tanh\left(\frac{\beta E}{2} \right).
\end{equation}
Were the two-level constraint imposed exactly, the result would instead be:
\begin{math}
  ( e^{\beta E} - e^{-\beta E})/(e^{\beta E} + e^{-\beta E})
  = \tanh\left(\beta E \right)
\end{math}, as the zero and doubly occupied states would be removed
from the denominator.} Note, that this is the standard mean-field gap
equation of the BCS-BEC crossover theory \cite{Randeria}.

\subsection{High temperature limit of Mean-field theory - simple laser}
\label{sec:mean-field-theory}

The opposite extreme to the equilibrium condensate is the limit of a
simple laser, which can also be recovered from Eq.~(\ref{eq:main}).
Before showing how this limit can be recovered from our theory, we
first provide a brief summary of the threshold condition of a simple
laser, and express it in similar language to the above
self-consistency condition.  The equations describing the steady state
of a laser can be derived starting from the well-known Maxwell-Bloch
equations:
\begin{eqnarray}
  \label{eq:33}
  \partial_t \psi_0 &=& - i \omega_0 \psi_0 - \kappa \psi_0 + \sum_i g_i P_i,
  \\
  \label{eq:34}
  \partial_t P_i &=& - 2 i \epsilon_i P - \lambda_{\perp} P_i + g_i \psi_0 N_i
  \\
  \label{eq:35}
  \partial_t N_i &=& \lambda_{\parallel}(N_0 - N_i) - 2 g_i (\psi_0^\ast P_i + P_i^\ast \psi_0).
\end{eqnarray}
These equations can be understood as originating from considering a
Hamiltonian like Eq.~(\ref{eq:1}), with $P_i=-i\langle a_i^\dagger b_i
\rangle, N_i=\langle b_i^\dagger b_i - a_i^\dagger a_i \rangle$. One
then writes the Heisenberg-Langevin equations, with a Markovian set of
baths distinct for each two-level system, and then takes the
semiclassical approximation to drop bath noise operators. The value
$N_0$ is the bath inversion imposed by the pumping.  Note that with
coupling to such a Markovian pumping bath, there is a discontinuous
jump between the allowed steady states with no decay, and the
laser-like solutions found for any non-zero
pumping\cite{szymanskaThesis}.  In particular, with pumping and decay,
inversion is always required for a condensed solution of these
Maxwell-Bloch equations, so they cannot smoothly interpolate between a
condensate and a laser.  Such behaviour should not be too surprising,
as a frequency independent (Markovian) bath occupation corresponds to
an infinite temperature, and so even arbitrarily weak coupling of the
system to an infinite temperature reservoir may destroy the
condensate.  With more realistic models of pumping, such a
discontinuous jump need not necessarily occur.  One should thus
interpret the microscopic origin and consequent behaviour of
Eqs.~(\ref{eq:33}--~\ref{eq:35}) with some caution. However, since
Maxwell-Bloch equations of the above form are frequently used as a
simple model of a laser, it is instructive to see what approximations
they would correspond to in terms of our non-equilibrium formalism, in
which the microscopic description of the pumping is better
controlled.

Starting from these Maxwell-Bloch equations, the self-consistency
condition for a macroscopic photon field $\psi_0(t) = \psi_0 e^{-i \mu
  t}$ can be written as:
\begin{equation}
  \label{eq:36}
  (-i \mu + i \omega_0 + \kappa) \psi_0 = \sum_i g_i P_i,
  \qquad
  (-i \mu + 2i \epsilon_i + \lambda_{\perp}) P_i = g_i \psi_0 N_i,
\end{equation}
which can be combined to write a single self-consistency condition:
\begin{equation}
  \label{eq:37}
  (\omega_0 - \mu - i \kappa) \psi_0 = - \sum_i
  \frac{g_i^2 \psi_0 N_i }{2 \tilde{\epsilon}_i - i \lambda_{\perp}}.
\end{equation}
Substituting the steady state value of $P_i$ from Eq.~(\ref{eq:36}) into
Eq.~(\ref{eq:35}) gives:
\begin{equation}
  \label{eq:38}
  N_0 
  =  N_i \left[
    1
    + \frac{2 g_i^2 |\psi_0|^2}{\lambda_{\parallel}}
    \frac{2 \lambda_{\perp} }{\lambda_{\perp}^2 + 4  \tilde{\epsilon}_i^2} 
  \right]
\end{equation}
hence we may substitute this into Eq.~(\ref{eq:37}) to give the final
form of the self-consistency condition for the Maxwell-Bloch
equations:
\begin{equation}
  \label{eq:39}
  (\omega_0 - \mu - i \kappa) \psi_0
  =
  - \sum_i  g_i^2 \psi_0  N_0
  \frac{  (2 \tilde{\epsilon}_i +  i\lambda_{\perp})}{
    [4 \tilde{\epsilon}_i^2 + \lambda_{\perp}^2 + 4 (\lambda_{\perp}/\lambda_{\parallel}) g_i^2 |\psi_0|^2]}.
\end{equation}
The laser threshold condition is given by taking $\psi \to 0$ in the
above equation.  If we also take $g_i=g$, $\epsilon_i=\epsilon$, and
the usual laser operating condition of $\lambda_{\perp} \gg \kappa$
one
has that lasing occurs at the cavity frequency, $\mu=\omega_0$ and the
threshold condition has the well-known form: $\kappa
\lambda_{\perp}/{g^2}=n N_0$, where $n$ is the number of two-level
systems.

\subsubsection{Recovering laser limit from non-equilibrium mean-field theory}
\label{sec:recov-laser-limit}

This simple laser self-consistency condition can be recovered from
equation~(\ref{eq:main}) if rather than using the frequency dependent
forms for $F_{A,B}(\nu)$ discussed previously, one instead takes
$F_{A,B}$ to be constants .  Physically such a limit corresponds to
high temperatures.  Note that as the temperature rises, to keep the
bath population fixed, the chemical potential must also vary.  We will
therefore take $\mu \propto T$, and then take the limit $T \to
\infty$.  Such a limit has another simple interpretation,
corresponding to making a fully Markovian approximation, including
assuming the occupation, as well as the density of states, to be flat,
and so writing the Keldysh part of the self energy as
\begin{equation}
  \label{eq:40}
  \Sigma_{a^\dagger a}^{--}(t,t^\prime) 
  = 
  - i\sum_n \Gamma_{i,n}^2 [1- 2n_F(\nu^\Gamma_n)] 
  e^{-i\nu^\Gamma_n(t-t^\prime)}
  =
  - 2 i \gamma F_A \delta(t-t^\prime).
\end{equation}
As such, our approach in Eq.~(\ref{eq:24}) is Markovian for the
density of states of the bath, but non-Markovian for the occupation.
In terms of quantum statistical (i.e. Heisenberg-Langevin) approaches,
the distinction is whether the noise should be taken as white noise or
coloured noise.  Assuming the noise correlations to be white, and thus
neglecting the frequency dependence of occupation, is also the
approximation underlying the quantum regression
theorem\cite{scully97}, which allows one to relate two-time
correlations to the evolution of the density matrix.  The role of this
approximation, and its implications for the fluctuation dissipation
theorem are discussed by Ford and O'Connel\cite{ford96}.

If $F_{A,B}$ are frequency independent, then in Eq.~(\ref{eq:main}), the
term in the integral proportional to $\nu$ will vanish as this is an
odd function, and so Eq.~(\ref{eq:main}) becomes:
\begin{equation}
  \label{eq:41}
  (\omega_0 - \mu - i \kappa) \psi_0
  =
  \sum_i g^2_i \psi_0  (F_B-F_A)
  \frac{\tilde{\epsilon}_i+i\gamma}{4(E_i^2+\gamma^2)}.
\end{equation}
Hence, the polarisation of the two-level systems is in this case
proportional to the inversion of the baths, $N_0 = (n_B-n_A) =
-(F_B-F_A)/2$ and we have:
\begin{equation}
  \label{eq:42}
  (\omega_0 - \mu - i \kappa) \psi_0  =
  - \sum_i g_i^2 \psi_0 N_0
  \frac{\tilde{\epsilon}_i+i\gamma}{2(E_i^2+\gamma^2)}.
\end{equation}
Then, identifying the decay constants in Eq.~(\ref{eq:39}) as
$\lambda_{\perp}=\lambda_{\parallel}=2 \gamma$, Eq.~(\ref{eq:42}) and
Eq.~(\ref{eq:39}) are equivalent.

\subsection{General properties of mean-field theory away from extremes}
\label{sec:mean-field-theory-1}

\begin{figure}[htpb]
  \centering
  \includegraphics[width=0.8\textwidth]{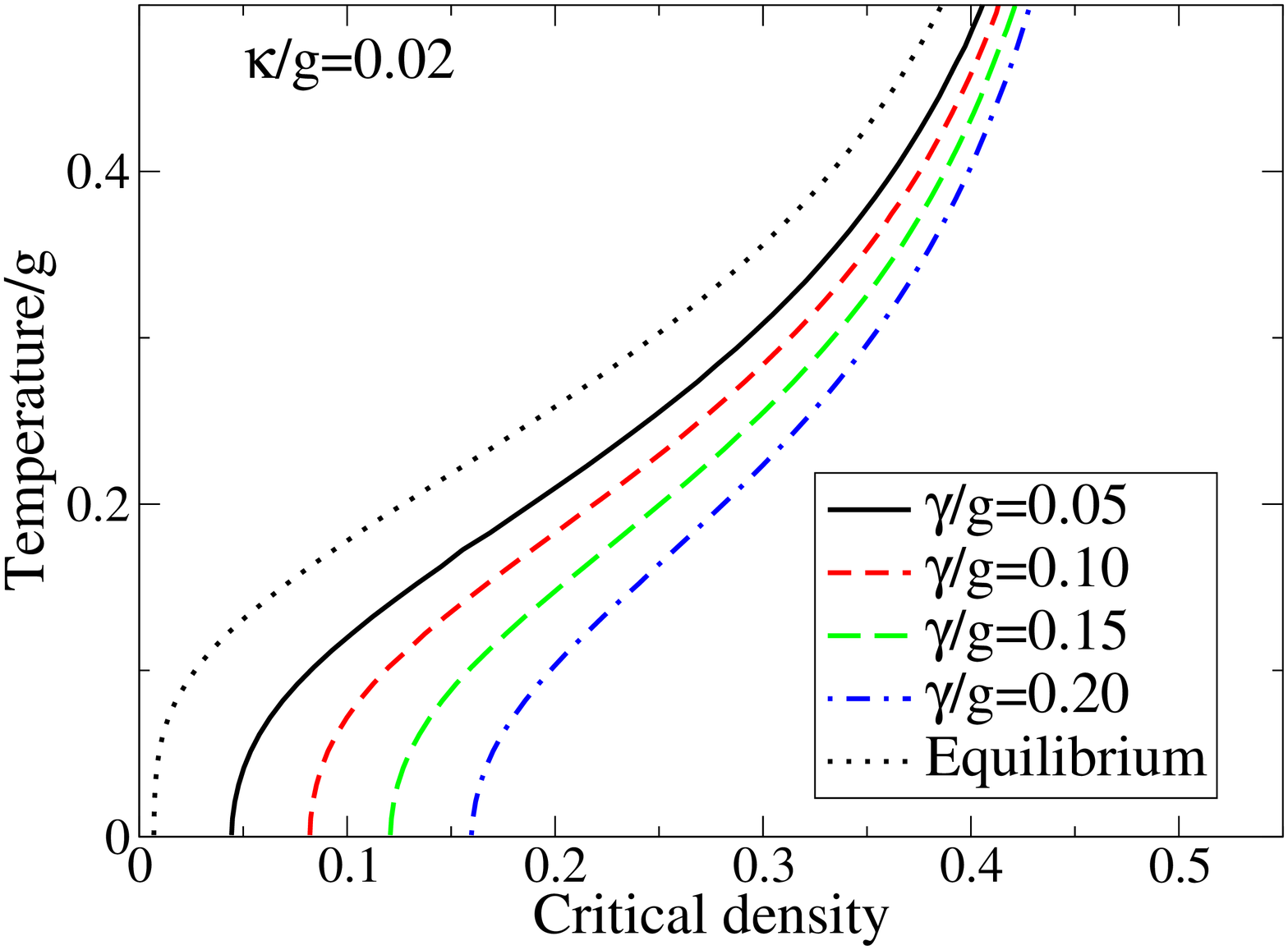}
  \caption[Critical temperature vs density]{Critical temperature as a
    function of density, showing effects of pumping and decay, taking
    a Gaussian distribution of two-level-system energies with variance
    $0.15 g$.  Adapted from Ref.\cite{szymanska07}.  }
  \label{fig:critical-temperature}
\end{figure}

Away from the extremes of laser theory or of thermal equilibrium, the
effect of pumping on the phase boundary can be understood as a result
of competition of two effects: pumping and decay add noise, reducing
coherence, hence suppressing condensation;  on the other hand, for a
given decay rate, pumping increases the density, favouring
condensation.  The simplest illustration of the first of these is
shown in Fig.~\ref{fig:critical-temperature}, where one sees that as
the value of $\gamma$ is increased, for a fixed $\kappa$, the critical
density required for condensation increases.

\begin{figure}[htpb]
  \centering
  \includegraphics[width=0.8\textwidth]{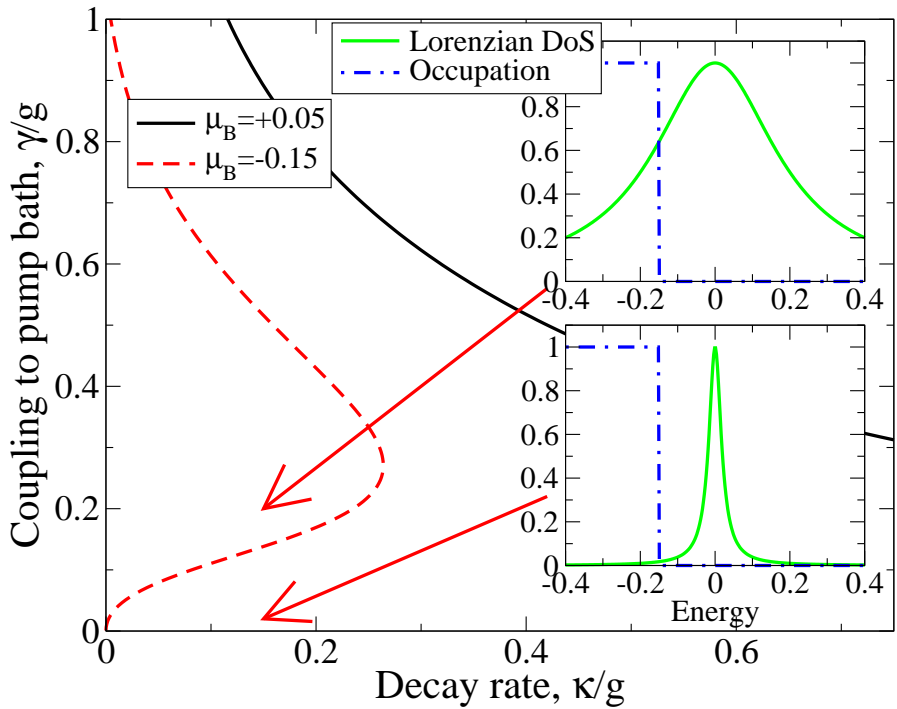}
  \caption[Critical $\kappa,\gamma$ for clean system.]{Critical
    couplings to pumping bath without inhomogeneous broadening and at
    low temperatures.}
  \label{fig:critical-kappa-gamma-clean}
\end{figure}

To see the competition between pumping causing dephasing and pumping
increasing density, one may look at the low temperature limit, shown
in Fig.~\ref{fig:critical-kappa-gamma-clean}, plotting the critical
value of $\kappa$ as a function of $\gamma$.  Two lines are shown; the
solid line has an inverted bath (as would be required for the laser
limit), the dashed line has a non-inverted bath.  In the later case
(as illustrated in the inset) for small $\gamma$, the two-level system
energy is too far below the pumping bath, and insufficiently broadened
by $\gamma$, to be populated; for larger $\gamma$ the broadening is
sufficient, and condensation may occur.  In the presence of
inhomogeneous broadening, the above picture is significantly relaxed,
since the tail of the density of states can be occupied even if the
peak is below the chemical potential.

\subsection{Low density limit: recovering complex Gross-Pitaevksii equation}
\label{sec:local-dens-appr}

The self-consistency condition of Eq.~(\ref{eq:main}) can also be
related to the idea of the complex Gross-Pitaevskii equation providing
a mean-field description of a spatially varying condensate.
For a steady uniform state, the mean-field self-consistency condition
may be understood as as $(\mu_S + i \kappa - \omega_0 ) \psi_0
= \chi[\psi_0, \mu_S] \psi_0$, where $\chi[\psi_0,\mu_S]$ is a
nonlinear complex susceptibility.  For a $\psi_0(r,t)$ which varies
slowly in space and time [up to an allowed fast time dependence
described by a factor $\exp(-i \mu_S t)$], one may consider the local
density approximation:
\begin{equation}
  \label{eq:43}
  \left(
    i \partial_t + i \kappa
    -
    \left[ V(r) - \frac{\nabla^2}{2m} \right]
  \right) \psi_0(r,t) = \chi[\psi_0(r,t)] \psi_0(r,t).
\end{equation}
In order to determine the large scale spatial structure of a
condensate, or its low energy collective modes, it is often sufficient
to make a Taylor expansion of the nonlinear complex susceptibility,
resulting in a complex Gross-Pitaevskii equation:
\begin{equation}
  \label{eq:44}
  i \partial_t
  \psi_0
  =
  \left(
    - \frac{\nabla^2}{2m}
    +
    V(r)  
    +
    U |\psi_0|^2
    +
    i \left[\gamma_{\mathrm{eff}}(\mu_B) -  \kappa -  \Gamma |\psi_0|^2 \right] 
  \right)
  \psi_0,
\end{equation}
where $\Gamma$ represents the simplest form of nonlinearity of the
imaginary part, taking a form that will ensure stability.

Depending on the details of pumping included in the model, one may
find that by treating $\chi[\psi(t)]$ more carefully the
susceptibility depends not only on the current value of $\psi(t)$, but
on its history, due to dynamics of the reservoir.  [In fact, to
correctly reproduce the polariton spectrum, one ought to take the
excitonic susceptibility to have a resonance at the exciton energy,
after which a variant of Eq.~(\ref{eq:44}), but with the appropriate
polariton dispersion will be recovered.]  In the limit of sufficiently
slow dynamics of the system, or when considering steady states,
dynamics of the reservoir should become unimportant.  Results of the
complex Gross-Pitaevskii equation with or without separate reservoir
dynamics may be found
elsewhere\cite{wouters07:bec,wouters08:bec,keeling08:gpe}

\section{Fluctuations, and instability of the normal state}
\label{sec:fluct-inst-norm}

As stated earlier, when introducing the self-consistency condition for
the non-equilibrium problem, it is not possible to consider minimising
free energy when looking at a system coupled to multiple baths, and so
it is instead necessary to look for stable steady states.  The self
consistency conditions discussed above determine whether a steady
state may exist, but not whether it is stable.  In order to analyse
stability, it is necessary to consider fluctuations about a given
state, and to find whether they grow or decay in time.  In addition,
the study of fluctuations allows one to determine the response
functions of the system --- in the present context, this means the
photon Green's function --- which in turn will give the physical
observables, such as photoluminescence and absorption spectra.

In the non-equilibrium case, both the spectrum of possible excitations
(i.e. what is seen in the absorption spectrum), and its occupation
(i.e.  photoluminescence) must be determined independently, for which
the Keldysh Green's function approach is ideal.  In the following, the
approach to calculating these Green's functions is discussed for both
the normal and condensed state, and then this approach is applied to
understanding the instability of the normal state, which allows a
clearer interpretation of the relation between the non-equilibrium
condensate and a simple laser.
For the condensed system, the calculations are more complicated due to
the existence of non-zero anomalous correlations, i.e. $\langle
\psi_k(t) \psi_{-k}(t^\prime) \rangle$; the general structure of the
spectrum of the non-equilibrium system will be discussed in
section~\ref{sec:fluct-cond-syst}.

\subsection{Photon Green's functions in the non-equilibrium model}
\label{sec:phot-greens-funct}

To allow for anomalous correlations in the condensed state, it is
helpful to write the Green's function in a vector space of $\psi_k,
\psi^\dagger_{-k}$.  Just as in the above discussion of the Green's
functions for the two-level system, this vector space of $\psi_k,
\psi^\dagger_{-k}$ should be combined with the $\pm$ space due to the
Keldysh/retarded/advanced structure.  Thus, one has four by four
matrices, in the basis $(\psi^\nodagger_{k,+}, \psi^\dagger_{-k,+},
\psi^\nodagger_{k,-}, \psi^{\dagger}_{-k,-})$.

The photon Green's function can be found by solving the
Dyson equation, ${D}^{-1}_{\vphantom{0}} = {D}^{-1}_0 - \Sigma$, and
so to start with, the free photon Green's function is required.  The
free Hamiltonian in this case is just $H^{{\mathrm{photon}}}_0 = \sum_k \omega_k
\psi^\dagger_k \psi_k$. In the four by four basis arising from mixing
$\psi_k, \psi^\dagger_{-k}$, some elements correspond to Green's
functions in which $\psi, \psi^\dagger$ are interchanged in order.
This means that these elements are Hermitian conjugated, giving the
form:
\begin{equation}
  \label{eq:45}
  {D}^{-1}_0 = \left(
    \begin{array}{cccc}
      0 & 0 & \omega - \tilde{\omega}_k - i \eta & 0 \\
      0 & 0 & 0 & -\omega - \tilde{\omega}_k + i \eta \\
      \omega - \tilde{\omega}_k + i \eta & 0 & (2 i \eta) F_0(\omega+\mu) & 0 \\
      0 & -\omega - \tilde{\omega}_k - i \eta  & 0 & (2 i \eta) F_0(-\omega+\mu) \\
    \end{array}
  \right),
\end{equation}
where once again $\eta$ is infinitesimal.
In this, we have written all frequencies measured relative to $\mu_S$,
meaning that we made the substitution $\psi_k \to e^{-i \mu_S t}(\psi_0
\delta_{k,0} + \psi_k)$.

To this free Green's function one must add self energies arising from
two parts of the interaction Hamiltonian.  The first is the coupling
between cavity photons and the decay bath; the second is the coupling
between the photons and the pumped two-level systems.  The first
contribution has a form exactly analogous to the coupling between the
two-level systems and pumping baths, i.e.:
\begin{eqnarray*} 
  \Sigma_{\psi^\dagger \psi}^{++}(t,t^\prime) &=&
  \sum_p \zeta_{p,k}^2 D^{--}_{\Psi^\dagger \Psi} 
  = 0
  \\
  \Sigma_{\psi^\dagger \psi}^{-+}(t,t^\prime) &=&
  \sum_p \zeta_{p,k}^2 D^{+-}_{\Psi^\dagger \Psi} 
  = 
  - i\sum_p \zeta_{p,k}^2 \theta(t-t^\prime) e^{-i\omega^\zeta_p(t-t^\prime)}
  =-i \kappa \delta(t-t^\prime)
  \\
  \Sigma_{\psi^\dagger \psi}^{+-}(t,t^\prime) &=& 
  \sum_p \zeta_{p,k}^2 D^{-+}_{\Psi^\dagger \Psi} 
  = 
  + i\sum_p \zeta_{p,k}^2 \theta(t^\prime-t) e^{+i\omega^\zeta_p(t-t^\prime)}
  =+i \kappa \delta(t-t^\prime)
  \\
  \Sigma_{\psi^\dagger \psi}^{--}(t,t^\prime) &=& 
  \sum_p \zeta_{p,k}^2 D^{++}_{\Psi^\dagger \Psi} 
  = 
  - i\sum_p \zeta_{p,k}^2 [2n_\Psi(\omega^\zeta_p)+1] 
  e^{-i\omega^\zeta_p(t-t^\prime)}
  =-2i \kappa \breve{F}_{\Psi}(t-t^\prime),
\end{eqnarray*}
where as before $\breve{F}_{\Psi}$ is the Fourier transform of the
$2n_\Psi(\omega)+1$, and the Markovian limit for the bath density of
states and coupling constant has been applied to get the final
expression; these terms thus give a self energy:
\begin{equation}
  \label{eq:46}
  \Sigma_{\mathrm{decay}}(\omega)  = \left(
    \begin{array}{cccc}
      0 & 0 &  + i \kappa & 0 \\
      0 & 0 & 0 & -i \kappa \\
      -i \kappa & 0 & -(2 i \kappa) F_\Psi(\omega+\mu_S) & 0 \\
      0 & + i \kappa  & 0 & -(2 i \kappa) F_\Psi(-\omega+\mu_S) 
    \end{array}
  \right).
\end{equation}

In calculating the self energy due to the coupling to two-level
systems, one may simplify the calculation by noting that only
$\Sigma^R_{\psi^\dagger\psi}, \Sigma^R_{\psi^\dagger\psi^\dagger},
\Sigma^K_{\psi^\dagger\psi}, \Sigma^K_{\psi^\dagger\psi^\dagger}$ are
independent; all other self energies can be related to these
quantities by Hermitian conjugation and/or swapping $\omega \to -
\omega$.  To generate the diagrams for these self energies,
we should first determine the interaction vertices that give
rise to such self energies.  
{
The relevant part of the interaction Hamiltonian here is the interaction
between two-level systems and incoherent photons, and so the relevant
contribution to $U$ comes from}
\begin{eqnarray}
  \label{eq:47}
  \int_C dt \tilde{H}^{{\mathrm{TLS-photon}}}_{\mathrm{int}} 
  &=&
  \int_{-\infty}^\infty dt
  g
  \left[ 
    \tilde{\psi}(t,f) \tilde{b}^\dagger_{i}(t,f) \tilde{a}^\nodagger_{i}(t,f) 
    -
    \tilde{\psi}(t,b) \tilde{b}^\dagger_{i}(t,b) \tilde{a}^\nodagger_{i}(t,b) 
    + \mathrm{H.c.}
  \right]
  \nonumber\\
   &=&
   \int_{-\infty}^\infty dt
   \frac{g}{\sqrt{2}}
   \left[ 
    \tilde{\psi}_+(t) 
    \left(
      \tilde{b}^\dagger_{i+}(t) \tilde{a}^\nodagger_{i-}(t) 
      +
      \tilde{b}^\dagger_{i-}(t) \tilde{a}^\nodagger_{i+}(t) 
    \right)
    \right.\nonumber\\&&\left.\qquad\quad\ \ \
      {}+
    \tilde{\psi}_-(t) 
    \left(
      \tilde{b}^\dagger_{i+}(t) \tilde{a}^\nodagger_{i+}(t) 
      +
      \tilde{b}^\dagger_{i-}(t) \tilde{a}^\nodagger_{i-}(t) 
    \right)
    + \mathrm{H.c.}
   \right].
\end{eqnarray}
The self energy diagrams thus consist of diagrams with one incoming
and one outgoing photon line, connected via the interaction vertices
in Eq.~(\ref{eq:47}), and the Green's functions for the two-level
system.  As is clear from Eq.~(\ref{eq:47}), the vertices all involve
the two-level system swapping between the $a$ and $b$ states.  Just as
for the diagrams describing the effects of the bath discussed in
Sec.~\ref{sec:effects-baths-system}, one must also keep track of the
$\pm$ labels on the fields.  To calculate,
for instance, the retarded self energy (i.e. the $-+$ component) 
it is clear that the vertices arising from the possible placements
of $\pm$ signs have the form:\\
\centerline{\includegraphics{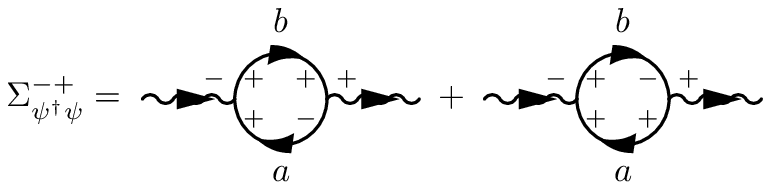}}\\
(any other set of possible $\pm$ labels on the internal lines will
involve a $--$ line, and such Green's functions vanish). 
To translate these diagrams into an equation for the self energy, one
must use the following Feynman rules (see
Refs.\cite{keldysh65,lifshitz:Phys_Kin,danielewicz84}): For each
interaction vertex there is a factor $(-i g/\sqrt{2})$, and for each
internal Green's function, a factor $i G$.  There is then a prefactor
$i (-1)^F$, where $F$ is the number of closed Fermion loops ($F=1$ in
the current case), and there is a combinatoric factor associated with
how the vertices are found from the expansion of $U$, which is the
same as in any other diagrammatic approach.  Applying these rules, one
may write:
\begin{equation}
  \label{eq:48}
  \Sigma^{-+}_{\psi^\dagger\psi}
  = -i\frac{2}{2!} \left(\frac{g}{\sqrt{2}}\right)^2
  \int \frac{d\nu}{2\pi}
  \sum_i 
  \left[
    G^{A}_{a^\dagger_i a^\nodagger_i}(\nu) G^K_{b^\dagger_i b^\nodagger_i}(\nu + \omega)
    +
    G^{K}_{a^\dagger_i a^\nodagger_i}(\nu) G^R_{b^\dagger_i b^\nodagger_i}(\nu + \omega)
  \right].
\end{equation}
For the anomalous case, all that changes is the $a,b$ labels, i.e.:
\begin{equation}
  \label{eq:49}
  \Sigma^{-+}_{\psi^\dagger\psi^\dagger}
  = -i\frac{2}{2!} \left(\frac{g}{\sqrt{2}}\right)^2
  \int \frac{d\nu}{2\pi}
  \sum_i 
  \left[
    G^{A}_{a_i^\dagger b^\nodagger_i}(\nu) G^K_{b_i^\dagger a_i^\nodagger}(\nu + \omega)
    +
    G^{K}_{a_i^\dagger b^\nodagger_i}(\nu) G^R_{b_i^\dagger a_i^\nodagger}(\nu + \omega)
  \right].
\end{equation}

The component $\Sigma^{+-}$ is just the Hermitian conjugate of
$\Sigma^{-+}$ as above. The component $\Sigma^{++}$ vanishes, since it
either involves $--$ lines, or it involves products of two retarded
Green's functions.  Since the retarded Green's function is causal ---
i.e.  $D^{R}(t,t^\prime) \propto \theta(t-t^\prime)$ --- then as a
function of frequency, all of its poles are in the lower half plane,
and so the integral of a product of two such functions is equal to
zero.
\footnote {NB; since the Green's function generically looks like
  $1/\omega$ at large $\omega$, the integral of a single retarded
  Green's function depends on the regularisation used.  However, for a
  product of retarded Green's functions, the integral is well defined,
  and so vanishes.}
The only other surviving component of the self energy is thus:\\
\centerline{\includegraphics{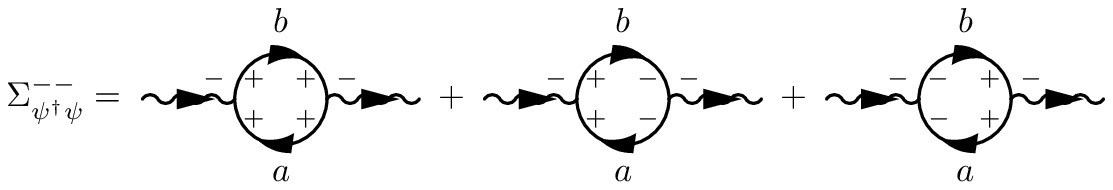}}\\
which gives the equation:
\begin{eqnarray}
  \label{eq:50}
  \Sigma^{--}_{\psi^\dagger\psi}
  &=& -i\frac{2}{2!} \left(\frac{g}{\sqrt{2}}\right)^2
  \int \frac{d\nu}{2\pi}
  \sum_i 
  \left[
    G^{K}_{a^\dagger_i a^\nodagger_i}(\nu) G^K_{b^\dagger_i b^\nodagger_i}(\nu + \omega)
    +
    G^{A}_{a^\dagger_i a^\nodagger_i}(\nu) G^R_{b^\dagger_i b^\nodagger_i}(\nu + \omega)
  \right.\nonumber\\&&\left.\qquad\qquad\qquad\qquad\qquad\quad
    {}+
    G^{R}_{a^\dagger_i a^\nodagger_i}(\nu) G^A_{b^\dagger_i b^\nodagger_i}(\nu + \omega)
  \right].
\end{eqnarray}
As was the case for the retarded components, the only difference
between normal and anomalous Keldysh components is in the $a,b$
labels, so:
\begin{eqnarray}
\label{eq:51}
  \Sigma^{--}_{\psi^\dagger\psi^\dagger}
  &=& -i\frac{2}{2!} \left(\frac{g}{\sqrt{2}}\right)^2
  \int \frac{d\nu}{2\pi}
  \sum_i 
  \left[
    G^{K}_{a_i^\dagger b^\nodagger_i}(\nu) G^K_{b_i^\dagger a_i^\nodagger}(\nu + \omega)
    +
    G^{A}_{a_i^\dagger b^\nodagger_i}(\nu) G^R_{b_i^\dagger a_i^\nodagger}(\nu + \omega)
  \right.\nonumber\\&&\left.\qquad\qquad\qquad\qquad\qquad\quad
    {}+
    G^{R}_{a_i^\dagger b^\nodagger_i}(\nu) G^A_{b_i^\dagger a_i^\nodagger}(\nu + \omega)
  \right].
\end{eqnarray}

Combining the self energies due to the pumped two-level systems
and the self energy due to decay with
the free inverse Green's function, one can then find expressions for
the photon Green's functions, and hence observable quantities such as
the photoluminescence intensity as a function of frequency and
momentum, which is given by $\mathcal{L}(\omega) = i \left(
  D^K_{\psi^\dagger\psi^\nodagger} - D^R_{\psi^\dagger\psi^\nodagger}
  + D^A_{\psi^\dagger\psi^\nodagger} \right)/2$.  In
section~\ref{sec:results-polar-model}, the normal state Green's
functions are studied: we show how an effective density of states and
occupation function can be defined, and also show how the behaviour of
these functions can be related to the structure of the inverse Green's
function, and to the stability of the normal system.

In the condensed state, just as in equilibrium, the
form of the inverse Green's function can be shown to
obey the Hugenholtz-Pines relation\cite{hugenholtz59} (see also
\cite[Chapter 6]{popov}), meaning that
$[D^{R}_{\psi^\dagger\psi}]^{-1}(0,0)
=[D^{R}_{\psi^\dagger\psi^\dagger}]^{-1}(0,0)$, which implies there is
a gapless spectrum.  Just as in equilibrium, one may show that the
requirement for the Hugenholtz-Pines relation to be satisfied is
equivalent to the mean-field condition, Eq.~(\ref{eq:main}).  It is
worth noting that as $\psi_0\to0$, the Hugenholtz-Pines relation (and
hence the mean-field condition) become equivalent to the condition
that:
\begin{equation}
  \label{eq:52}
  [D^{R}_{\psi^\dagger\psi}]^{-1}(\omega=\mu_{\mathrm{eff}} ,k=0) =
  \mu_{\mathrm{eff}} - \omega_0 + i \kappa - \Sigma^R_{\mathrm{TLS}}(\mu_{\mathrm{eff}}) = 0 ,
\end{equation}
for some particular $\mu_{\mathrm{eff}}$. (In this expression, the
non-condensed self energies have been written without the gauge
transform of Eq.~(\ref{eq:18}), as in the absence of a condensate,
there is no reason to perform the gauge transformation.)
Section~\ref{sec:results-polar-model} will show that the condition in
Eq.~(\ref{eq:52}) also corresponds to the point when the normal state
ceases to be stable.

\subsection{Normal-state Green's functions and instability}
\label{sec:results-polar-model}

Focusing on the non-condensed case, the properties of the spectrum
are entirely determined by three real functions of $\omega$, as one may write:
\begin{equation}
  \label{eq:53}
  \left[D_{\psi^\dagger\psi}^R\right]^{-1}(\omega) = A(\omega) + i B(\omega),
  \qquad
  \left[D_{\psi^\dagger\psi}^{-1}\right]^K(\omega)=i C(\omega).
\end{equation}
The forms of $A(\omega), B(\omega), C(\omega)$ follow from the
expressions in the previous section.  These somewhat simplify since
we are considering the normal case, and so one has:
\begin{equation}
  \label{eq:54}
  \left[D_{\psi^\dagger\psi}^R\right]^{-1}(\omega) = \omega - \omega_k + i \kappa
  - \Sigma^R_{\mathrm{TLS}}(\omega), \quad
  \left[D_{\psi^\dagger\psi}^{-1}\right]^K(\omega) = 2 i \kappa F_{\Psi}(\omega)
  -\Sigma^K_{\mathrm{TLS}}(\omega),
\end{equation}
where $\Sigma^{R,K}_{\mathrm{TLS}}$ are the self energies from the
pumped two-level systems given by Eq.~(\ref{eq:48}) and
Eq.~(\ref{eq:50}).  In the following we will first discuss how the
forms of $A(\omega), B(\omega), C(\omega)$ determine the spectrum,
occupation and stability, and then illustrate this with their forms
arising from the particular microscopic model discussed above.

Inverting the matrix of Keldysh Green's functions (using
Eq.~(\ref{eq:8})), one finds:
\begin{equation}
  \label{eq:55}
  D_{\psi^\dagger\psi}^R(\omega) = \frac{A(\omega)-i B(\omega)}{A(\omega)^2+B(\omega)^2},
  \qquad
  D_{\psi^\dagger\psi}^K(\omega) = \frac{-i C(\omega)}{A(\omega)^2+B(\omega)^2},
\end{equation}
and then in terms of these quantities, we may write the luminescence spectrum:
\begin{equation}
  \label{eq:56}
  \mathcal{L}(\omega) 
  =
  \frac{i}{2} \left[ 
    D^K_{\psi^\dagger\psi}(\omega)
    -
    \left(
    D^R_{\psi^\dagger\psi}(\omega)
    -
    D^A_{\psi^\dagger\psi}(\omega)
    \right)
  \right]
  =
  \frac{C(\omega) - 2 B(\omega)}{%
    2[A(\omega)^2+B(\omega)^2]}. 
\end{equation}
Further, by analogy with the equilibrium system, we can explain the
form of this expression in terms of a spectral weight (density of
states) $\rho(\omega) = -2 \Im[D_{\psi^\dagger\psi}^R(\omega)]$ and an
occupation function $2n_{\psi}(\omega)+1 = i
D_{\psi^\dagger\psi}^K(\omega)/\rho(\omega)$, giving:
\begin{equation}
  \label{eq:57}
  \rho(\omega) = \frac{ 2B(\omega)}{A(\omega)^2+B(\omega)^2},
  \qquad
  n_\psi(\omega) = \frac{1}{2}\left[ 
    \frac{C(\omega)}{2B(\omega)}
    - 1
  \right],
\end{equation}
hence the luminescence is related to these as $\mathcal{L}(\omega) =
\rho(\omega) n_\psi(\omega)$ as expected.

In the absence of coupling to the two-level systems (and hence
neglecting $\Sigma^{R,K}_{\mathrm{TLS}}$ in Eq.~(\ref{eq:54})
, one may clearly identify
the role of the three expressions involved here:
\begin{itemize}
\item  $B(\omega) = \kappa$ is the linewidth of the normal
  modes
\item $A(\omega)= \omega-\omega_k$ describes the locations
  of these modes, and
\item $C(\omega) = 2 \kappa (2 n_\psi + 1)$ describes their occupation.
\end{itemize}
However, when coupled to the two-level systems, $B(\omega)$ is not a
constant, hence firstly, the linewidth varies, and more importantly,
$B(\omega)$ may vanish at some value of $\omega$. If $B(\omega)$ does
vanish then the occupation diverges, but since the spectral weight
vanishes too, the luminescence remains finite.

Physically, this describes the behaviour that would, in equilibrium,
be expected at the chemical potential, as long as the chemical
potential lies below the bottom of the band. Note that the equilibrium
Bose-Einstein distribution diverges at the chemical
potential. However, if the chemical potential lies below the bottom of
the band then the spectral weight is zero at the chemical potential
and thus the particle number (luminescence) remains finite.

\begin{figure}[htpb]
  \centering
  \includegraphics[width=3.5in]{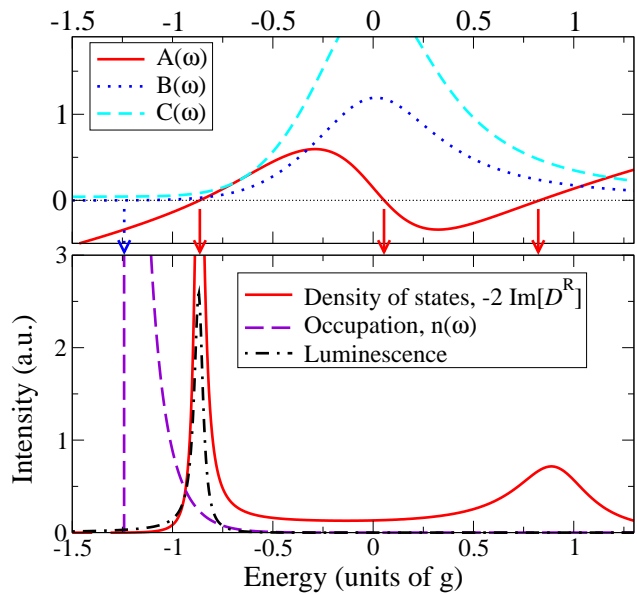}
  \caption[Inverse Green's functions and spectral weight]{Behaviour of
    inverse Green's functions, and resulting properties of spectral
    weight, luminescence and occupation functions in the normal state.
    Upper panel shows the inverse Green's functions (with zeros marked
    by arrows), and the lower panel shows the various physical
    correlations of interest.  Adapted from Ref.\cite{szymanska07}.}
  \label{fig:normal-luminescence}
\end{figure}

Out of equilibrium, the system distribution may in general be far from
the Bose-Einstein distribution.  Even so, when near the threshold for
condensation, the system distribution shares an important property
with the Bose-Einstein distribution: near the frequency where the
condensate will emerge [i.e near the point where $B(\omega)=0$] the
system distribution will diverge as $1/(\omega-\mu_{\mathrm{eff}})$,
just as the Bose-Einstein distribution does.
We may thus identify the effect of pumping as introducing a chemical
potential that has nothing to do with the chemical potential of the
decay bath.  Since $B(\omega)$ is given by the inverse retarded
  Green's function, one may note that the inverse Keldysh
Green's function does not on its own fix the
distribution; it is the ratio of Keldysh and imaginary retarded
Green's functions that matter.  Figure~\ref{fig:normal-luminescence}
shows how the spectral weight, occupation and luminescence are related
to the zeros of the real and imaginary parts of the inverse Green's
function.

\subsubsection{Zeros of $A(\omega), B(\omega)$ and stability}
\label{sec:zeros-aomega-bomega}

Although a zero of $B(\omega)$ alone does not cause the luminescence
to diverge, a simultaneous zero of  $A(\omega)$ and $B(\omega)$
will.  The stability of the system can be seen to change when this
occurs, as will be discussed next.  When near a simultaneous zero, one
may expand $A(\omega) = \alpha(\omega - \xi)$, and $B(\omega) = \beta(\omega
- \mu_{\mathrm{eff}})$, and so:
\begin{eqnarray}
  \label{eq:58}
  [D_{\psi^\dagger\psi}^R]^{-1}(\omega)
  &\simeq& \alpha(\omega - \xi) + i \beta(\omega-\mu_{\mathrm{eff}})
  \nonumber\\&=&
  (\alpha+i \beta) \left[
    \omega - \frac{(\alpha \xi + i \beta \mu_{\mathrm{eff}})(\alpha-i\beta)}{%
      \alpha^2+\beta^2}
  \right],
\end{eqnarray}
hence the actual poles are at frequencies:
\begin{equation}
  \label{eq:59}
  \omega^\ast = 
  \frac{%
    (\alpha^2 \xi + \beta^2 \mu_{\mathrm{eff}}) + i \alpha\beta (\mu_{\mathrm{eff}} - \xi)}{%
    \alpha^2+\beta^2}.
\end{equation}
These poles determine the time dependence of the retarded Green's
function, so if $\mu_{\mathrm{eff}} > \xi$, then the pole has the
wrong sign of imaginary part and the normal state is unstable.  When
$\mu_{\mathrm{eff}} = \xi$, then this means there is a value
$\omega=\mu_{\mathrm{eff}} = \xi$ for which
$\left[D_{\psi^\dagger\psi}^R\right]^{-1}(\omega,k=0) = 0$,
which as discussed in Eq.~(\ref{eq:52}) is equivalent to saying that
the mean-field consistency condition can be satisfied. Hence,
instability of the normal state, and the existence of a condensed
solution will occur together.

It is helpful here to explicitly write $B(\omega)$, in order
to understand the origin of its zeros, and what parameters determine
their location.  In the non-condensed case, the fermionic Green's
functions that come from inverting Eq.~(\ref{eq:25}) have a simple
form:
\begin{equation}
  \label{eq:60}
  G^R_{b^\dagger b, a^\dagger a}= \frac{1}{\nu \mp \epsilon_i + i \gamma},
  \qquad
  G^K_{b^\dagger b, a^\dagger a}= - \frac{2 i \gamma F_{B,A}(\nu)}{
    (\nu \mp \epsilon_i)^2 + \gamma^2},
\end{equation}
and so substituting these into Eq.~(\ref{eq:48}), and taking the imaginary
part one may write:
\begin{equation}
  \label{eq:61}
  B(\omega) 
  =
  \kappa +
  \gamma^2
  \int \frac{d\nu}{2\pi} \sum_i g_i^2
  \frac{  F_B(\nu + \omega) -   F_A(\nu)}{%
    \left[\left(\nu+\omega-\epsilon_i\right)^2 + \gamma^2\right]
    \left[\left(\nu+\epsilon_i\right)^2 + \gamma^2\right]
  }.
\end{equation}
For $B(\omega)$ to have zeros, it is necessary that the second term
(which originates from pumping) should be negative,
and should overcome the first term (which originates from
decay).  With $F_{A,B}(\nu) = \tanh\left[ \beta (\nu \pm
  \mu_B/2)/2\right]$, it is clear that this criterion requires $\mu_B$
to be sufficiently large.  As such, the following scenario describes
what happens as $\mu_B$ is increased:
\begin{description}
\item[Very weak pumping.] For large negative $\mu_B$, one finds that
  $F_{B}(\nu + \omega) - F_A(\nu)$ is always positive, and
  so no zero of $B(\omega)$ exists.
\item[Subcritical pumping.] For less negative values of $\mu_B$, there
  is a range of $\omega$ for which $B(\omega)$ is negative, indicating
  a range of gain in the spectrum.  The boundary of this region, where
  $B(\omega)=0$ defines an effective chemical potential
  $\mu_{\mathrm{eff}}$, but since $\mu_{\mathrm{eff}} < \xi$ the
  normal state remains stable.
\item[Critical pumping.] At some value of $\mu_B$, one finds that
  $\mu_{\mathrm{eff}}=\xi$, meaning that at this value of
  $\omega^\ast=\mu_{\mathrm{eff}}=\xi$, one has $D_{\psi^\dagger\psi}^{R}(\omega^\ast)
  =0$.  Hence, the gap equation first has a solution at this point,
  there is a real divergence of the luminescence, and the normal state
  is marginally stable.  
\item[Supercritical pumping.] Above this critical value of $\mu_B$,
  the normal state would have $\mu_{\mathrm{eff}} > \xi$, and so would
  be unstable.
\end{description}

\begin{figure}[htpb]
  \centering
  \includegraphics[width=3.5in]{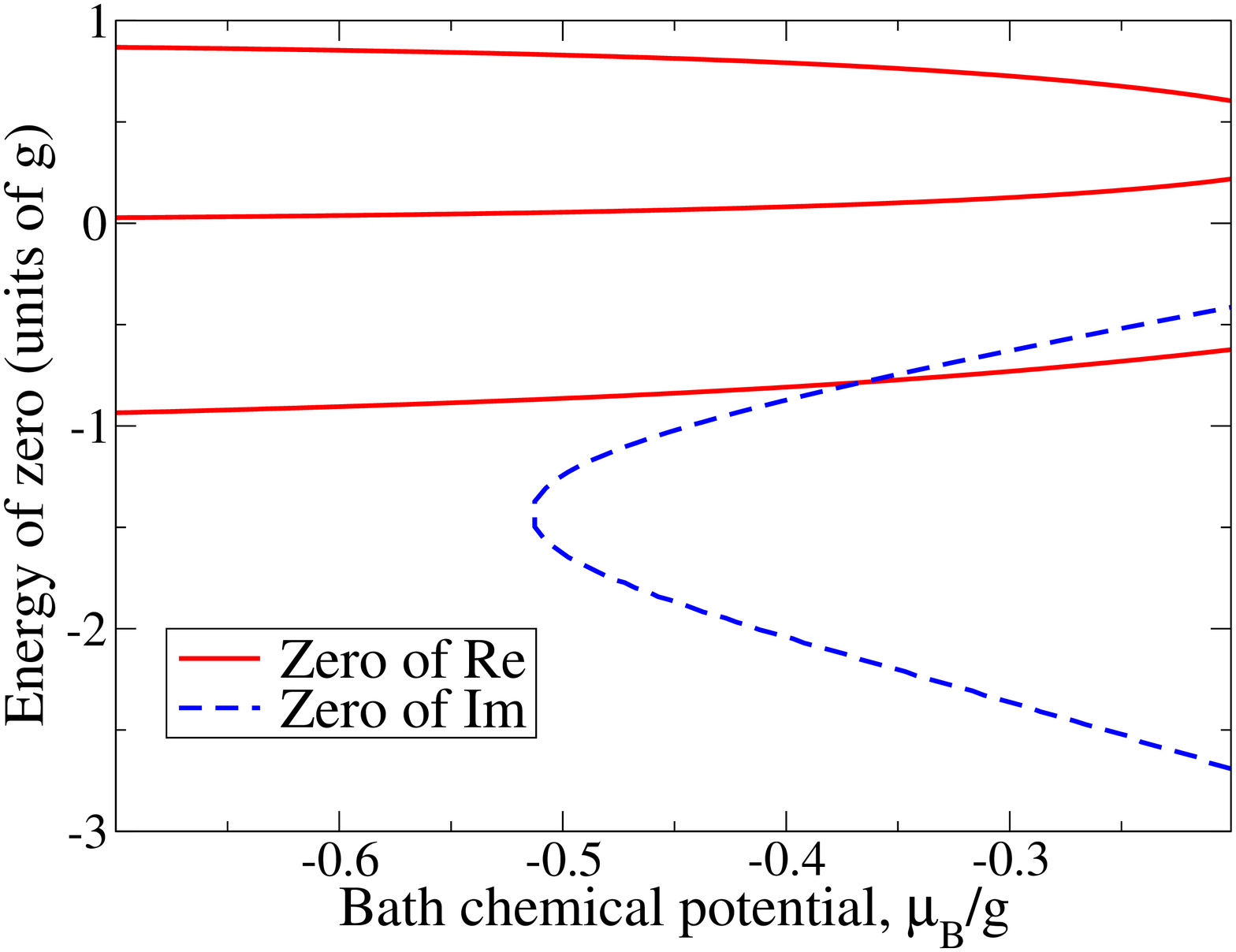}
  \caption[Zeros of {$\Re,\Im\left[(D_{\psi^\dagger\psi}^R)^{-1}\right]$} from
  diagrammatic approach.]{Variation of energies of zeros of real part
    of inverse Green's function and imaginary part as density is
    varied via chemical potential of pumping bath. The three solid
  lines correspond to (from the bottom) lower polariton, exciton, and
  upper polariton respectively. The point where the dashed line crosses
  solid line is where the condensation occurs.
 Adapted from
    Ref.\cite{szymanska07}.}
  \label{fig:normal-pole-traces}
\end{figure}

The actual behaviour for the polariton model of Eq.~(\ref{eq:1}) is
shown in Fig.~\ref{fig:normal-pole-traces}; one can see that a pair of
zeros of the imaginary part emerge, and then one crosses the bottom of
the polariton modes.  Note that in equilibrium, we have
$\mu_{\mathrm{eff}}=\mu=\mu_B$ at all conditions, and so only the last
three stages of the above scenario exist; condensation occurs when the
chemical potential reaches the bottom of the band.  It is also
important to note that the above scenario means that the Bose-Einstein
distribution is not the only distribution that would allow
condensation.  Any distribution which has the above property, i.e. a
divergence at some frequency for given values of the control
parameters (density, coupling constant, etc.), is sufficient for
quantum condensation in bosonic systems.

\subsubsection{Simplified form of distribution function in
  high-temperature limit}
\label{sec:simpl-dr-high}

\begin{figure}[htpb]
  \centering
  \includegraphics[width=3.5in]{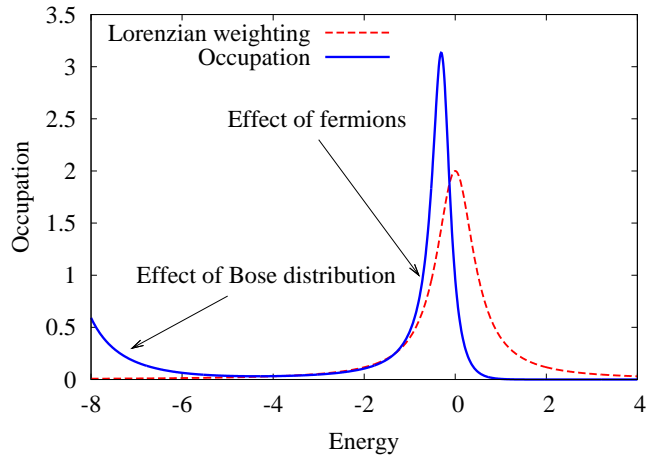}
  \caption[Distribution set by competition of baths]{Cartoon of
    occupation function set by competition of bosonic bath and
    fermionic bath, with effect of fermionic bath moderated by a
    Lorentzian filter depending on excitonic energy.  The chemical
    potential of the decay bath is at $-9$, and that of the pumping
    bath is just below zero.}
  \label{fig:partial-distributions}
\end{figure}

The way in which the effective distribution is set by the balance of
pumping and decay can be demonstrated more clearly by specialising to
the case of $\gamma \ll T$, for which the pumping bath occupation
functions do not change significantly across each Lorentzian broadened
peak (but may vary between the two peaks).  In
addition, consider taking $g_i=g, \epsilon_i=\epsilon$, so that
sums of two-level systems can be replaced by factors $n$. Then
the expression in Eq.~(\ref{eq:61}) can be simplified to give:
\begin{equation}
  \label{eq:62}
  B(\omega) 
  =
  \kappa +
   n g^2 \gamma
  \frac{[F_B(\epsilon) -   F_A(\epsilon-\omega)]}{%
    \left(\omega-2\epsilon\right)^2 + 4 \gamma^2},
\end{equation}
where we have performed the integrals assuming the distributions are
effectively constant.
\footnote{ Formally, the approximation consists
  of performing the contour integral, taking into account the poles at
  $\nu = -\omega + \epsilon_i + i \gamma$ and $\nu=-\epsilon+
  i\gamma$, but neglecting the poles from $F_{A,B}(\nu)$ which are at $\nu =
  \{-\omega + \mu_B/2,-\mu_B/2\} + i ( 2n+1)\pi T$, along with
  neglecting $\beta \gamma$ in evaluating the residues.}  By applying
the same approach to $[D_{\psi^\dagger\psi}^{-1}]^K$ one has:
\begin{equation}
  \label{eq:63}
  2 n_\psi(\omega) + 1 = 
  \frac{%
    \kappa (2 n_\Psi(\omega) + 1)
    + \displaystyle
    \frac{n g^2 \gamma}{(\omega-2\epsilon)^2 + 4 \gamma^2}{%
      [1 - F_B(\epsilon)F_A(\epsilon-\omega)]}}{%
    \kappa \phantom{(2 n_\Psi(\omega) + 1)}
    + \displaystyle
    \frac{n g^2 \gamma}{(\omega-2\epsilon)^2 + 4 \gamma^2}{%
      [F_B(\epsilon)-F_A(\epsilon-\omega)]}}.
\end{equation}
>From this expression one may first note that if $\gamma=0$ (or more
generally if $\kappa \gg g^2 \gamma /[(\omega-2\epsilon)^2 + 4
\gamma^2]$), the the system distribution is the same as the
distribution of the decay bath (the photons outside the cavity) and so
$n_\psi(\omega) = n_\Psi(\omega)$.  On the other hand, if $\kappa=0$,
(or more generally, if $\kappa \ll g^2 \gamma /[(\omega-2\epsilon)^2 +
4 \gamma^2]$, which can occur near $\omega=2\epsilon$), the
distribution is set by the pumping bath.  In this case, the important
terms in Eq.~(\ref{eq:63}) are:
\begin{equation}
  \label{eq:64}
  2 n_\psi(\omega) + 1 
  =
  \frac{%
    1 - F_B(\epsilon)F_A(\epsilon-\omega)}{%
    F_B(\epsilon)-F_A(\epsilon-\omega)}
  =
  \coth\left(
    \frac{\beta}{2}
    \left[\epsilon-\frac{\mu_B}{2}
      -\epsilon+\omega-\frac{\mu_B}{2}\right]\right),
\end{equation}
which is a Bose distribution with the temperature and chemical
potential of the pumping bath.  Thus, the photon distribution
interpolates between the decay and pumping bath, depending on the
efficiency of coupling as a function of energy.  An illustration of
how this might look when the chemical potential of the decay bath is
not too dissimilar from the pumping bath is shown in
Fig.~\ref{fig:partial-distributions}, however for realistic parameters, the
chemical potential of the decay bath should be taken to $-\infty$.

\subsection{Normal-state instability for a simple laser }
\label{sec:norm-state-inst}

As for the mean-field theory, it is instructive to compare the results
of Sec.~\ref{sec:zeros-aomega-bomega} to those for a simple laser, in
which pumping tries to fix the inversion of the gain medium,
independent of frequency.  The instability of the normal state can
still be determined by the inverse retarded Green's function, which
may in turn be found by the response of
Eq.~(\ref{eq:33})--(\ref{eq:35}) to an applied force $F e^{-i\omega
  t}$ acting on the photons.  If the force is weak, then
Eq.~(\ref{eq:35}) reduces to $N_i=N_0$, and taking $\lambda_{\perp}=2
\gamma$ as found previously, the equations to solve are:
\begin{equation}
  \partial_t \psi = - i \omega_0 \psi - \kappa \psi + \sum_i g_i P_i + F e^{-i\omega t},
  \qquad
  \partial_t P_i = - 2 i \epsilon_i P_i - 2 \gamma P_i + g \psi N_0,
\end{equation}
hence writing the response as $\psi=iD_{\psi^\dagger\psi}^R(\omega) F e^{-i\omega t}$, and
eliminating $P_i$ gives:
\begin{equation}
  \label{eq:65}
  [D_{\psi^\dagger\psi}^R]^{-1}(\omega) = \omega - \omega_0 + i \kappa 
  + \sum_i\frac{g_i^2 N_0}{\omega - 2  \epsilon_i + i 2 \gamma}.
\end{equation}
As in the mean-field case, this same equation can be recovered from
the microscopic non-equilibrium model by taking $F_{A,B}$ to
be independent of frequency, and identifying $N_0 = - (F_B - F_A)/2$.
The form of the inverse retarded Green's function makes much clearer
the implications of this absence of frequency dependence.  For the
imaginary part of Eq.~(\ref{eq:65}) to be zero, it is clearly
necessary that $N_0 > 0$, so a region of gain can only exist when
inverted.

\begin{figure}[htpb]
  \centering
  \includegraphics[width=3.5in]{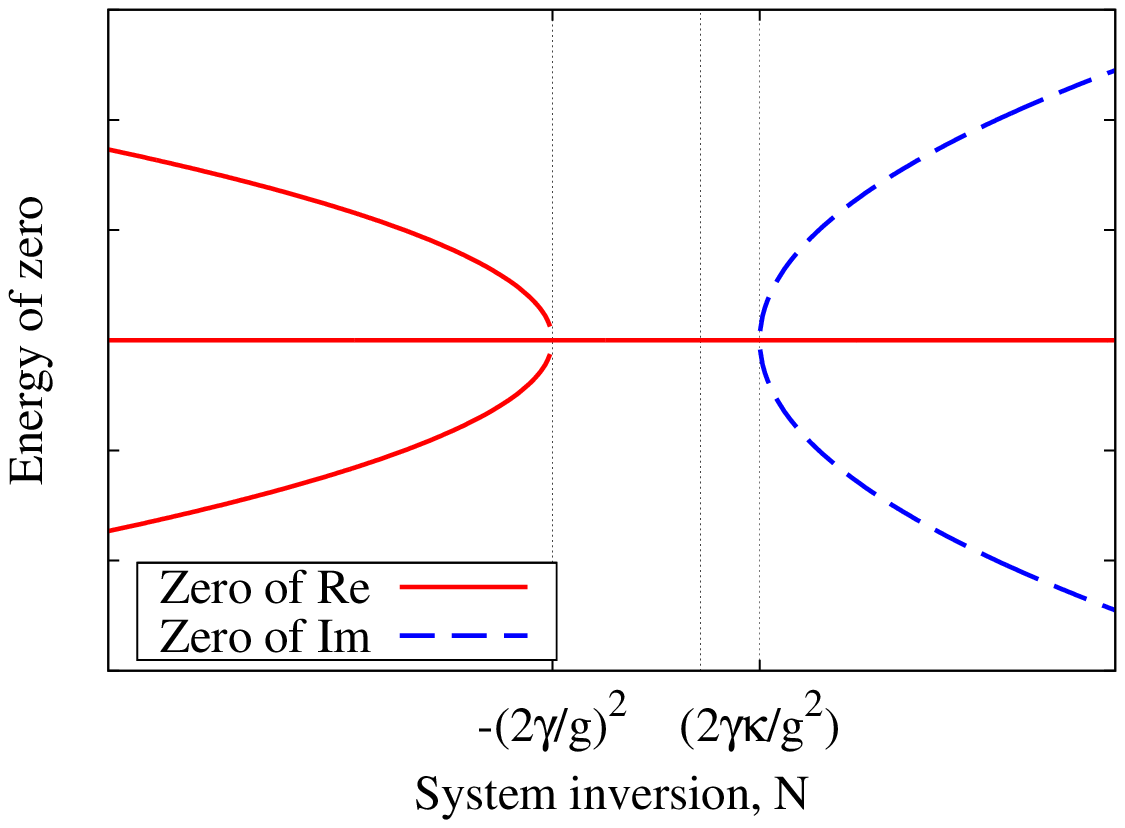}
  \caption[Zeros of {$\Re,\Im\left[(D_{\psi^\dagger\psi}^R)^{-1}\right]$} from
  Maxwell-Bloch equations.]{As for Fig.~\ref{fig:normal-pole-traces}
    but for the results of the Maxwell-Bloch equations, showing the
    rather different behaviour in the extreme laser limit, plotted
    for $\omega_0=2\epsilon$.}
  \label{fig:wc-laser}
\end{figure}

In the special case of $\epsilon_i = \epsilon=\omega_0/2, g_i=g$,
the zeros of the real and imaginary parts can be found explicitly to be
\begin{equation}
  \label{eq:66}
  \mu_{\mathrm{eff}} = 
  2 \epsilon \pm \sqrt{ g^2 n N_0 \frac{2 \gamma}{\kappa} - 4 \gamma^2}.
  , \qquad
  \xi = 2 \epsilon,\quad 2\epsilon \pm \sqrt{- 4 \gamma^2 - g^2 N_0 n},
\end{equation}
where $n$ is the number of two-level systems as before.  From the
zeros of the imaginary part, one sees that a region of gain exists
only for $N_0 > 2 \kappa \gamma/g^2n$ (note that this is the laser
threshold condition discussed in section \ref{sec:mean-field-theory}).
On the other hand, a splitting of the zeros of the real part $\xi$
exists only if $N_0 < -4 \gamma^2/g^2{n}$. Thus the instability of the
normal state only occurs after the normal mode splitting has
collapsed.  This is illustrated in Fig.~\ref{fig:wc-laser}.  In this
figure, it is also clear that as soon as there is a region of gain,
there is an instability.  This is quite different from
Fig.~\ref{fig:normal-pole-traces}, where a region of gain, and thus
zeros of the imaginary part, emerged at a lower pumping strength than
was required for the instability.  This meant that in the
non-equilibrium condensate, a diverging distribution function exists
before condensation occurs, whereas for Fig.~\ref{fig:wc-laser}, the
distribution function has no divergence in the normal state.
\footnote{If one considers the more general case with detuning,
  $\omega_0 \neq 2 \epsilon$, a region of gain may appear before the
  instability occurs.  Furthermore, if one also has inhomogeneous
  broadening, $\epsilon \neq \epsilon_j$, and different inversion
  for different two-level systems, a region of gain can coexist with a
  splitting of the normal states.  However, the results for the
  non-equilibrium condensate shown in
  Fig.~\ref{fig:normal-pole-traces} had neither detuning nor
  inhomogeneous broadening; hence in the absence of such
  complications, the difference between the non-equilibrium condensate
  and a simple laser are particularly obvious.}

\section{Fluctuations of the condensed system}
\label{sec:fluct-cond-syst}

When condensed, the derivation of the spectrum from the inverse
Green's function written previously becomes much more involved, but
the essential features of the spectrum can be determined by
considering the symmetries that the system must possess --- the
results of this analysis are confirmed by the exact expressions for
the inverse Green's functions.  In particular, one may combine the
Hugenholtz-Pines relation, mentioned at the end of
Sec.~\ref{sec:phot-greens-funct}, with the analytic properties of the
Green's functions which imply
$[D^{R}_{\psi\psi^\dagger}]^{-1}(\omega,p) =
[D^{R}_{\psi^\dagger\psi}]^{-1}(-\omega,p)^{\ast}$,
$[D^{R}_{\psi^\dagger\psi^\dagger}]^{-1}(\omega,p) =
[D^{R}_{\psi\psi}]^{-1}(-\omega,p)^{\ast}$.
%, and
>From these general considerations, one may find that
the most general structure for sufficiently small $\omega, k$ is:
\begin{equation}
  \label{eq:67}
  D^R_{\psi^\dagger\psi}(\omega,k) = \frac{C}{\mathrm{det}([D^R]^{-1})}
  =
  \frac{C}{\omega^2 + 2 i \omega x - c^2 k^2},
\end{equation}
where $x$ is an effective linewidth, and $c$ an effective sound
velocity.  The form of this expression is dictated by: the need to
combine symmetry under $k\to -k$; the existence of a finite linewidth;
and the pole at $\omega=0, k=0$ that is ensured by the
Hugenholtz-Pines relation.  Higher order contributions could exist
(and in fact do exist) for larger $\omega, k$, but the
$\omega,k \to 0$ structure is fixed by these considerations.

The above structure means that the poles of the Green's function for
small $k$ are diffusive, i.e. $\omega^\ast = - i x \pm i \sqrt{ x^2 -
  c^2 k^2}$, meaning that long wavelength excitations decay, but with
a lifetime that diverges for one mode as $k\to 0$.  This same form is
also recovered from other approaches to non-equilibrium
condensates\cite{wouters07:bec}, including also the case of a
parametrically pumped polariton system\cite{wouters07:opo}.  Note that
if one were to naively extract a Landau critical velocity from the
real part of $\omega^\ast$, then this critical velocity would vanish.
There has been some work on how the concept of the Landau critical
velocity may be generalised for parametrically pumped
condensates\cite{carusotto04,ciuti05,amo09}, however the full
implications of the diffusive structure on superfluidity of
incoherently pumped non-equilibrium condensates remains an open
question.  Because the polariton system is two-dimensional, phase
fluctuations can be expected to play a particularly important role,
therefore the remainder of this section will discuss how the above
form of the Green's function determines the long-time correlations,
and hence the lineshape, and how this connects to other approaches to
deriving the polariton lineshape.

To take full account of the phase fluctuations, one must
reparameterise the fluctuations as $\psi = \sqrt{\rho+\pi}e^{i \phi}$.
In order that one works with fields for which there is a macroscopic
expectation of $\langle \psi \rangle$ this reparameterisation must be
performed in real space, and in terms of the fields on the forward and
backward contours, rather than the symmetric and antisymmetric
combinations of these fields. (Note that the macroscopic expectation
of the anti-symmetric combination $\psi_-$ vanishes~\cite{kamenev05}.)
To describe the long-time correlations, we wish to find the first
order coherence function ${D}^{fb}_{\psi^\dagger\psi}(t) = -i \langle
T_c [ \psi(t,f) \psi^\dagger(0,b)] \rangle$, {and} corresponds to the
Fourier transform of the luminescence spectrum, $\mathcal{L}(\omega)$.
Since it is the phase fluctuations that dominate the long time
behaviour, one may write this asymptotic behaviour in the form:
\begin{equation}
  \label{eq:68}
  {D}^{fb}_{\psi^\dagger\psi}(t)
  \simeq 
  \rho_{QC} \left<
    \exp\left[ i \left( \phi(t) - \phi(0) \right) \right] \right>
  =
  \rho_{QC} \exp[- f(t)],
\end{equation}
where $\rho_{QC}$ is the quasi-condensate density.  The function
$f(t)$ is given by the phase-phase correlation functions, and in two
dimensions is given by:
\begin{equation}
  \label{eq:69}
  f(t) = i \left[D^{fb}_{\phi\phi}(t) - D^{fb}_{\phi\phi}(0) \right]
  =
  \int \frac{d \omega}{2\pi} \int \frac{k dk}{2\pi}
  \left[ 1 - e^{-i\omega t} \right]
  i D^{fb}_{\phi\phi}(\omega, k).
\end{equation}
Note that expressions (\ref{eq:68}) and (\ref{eq:69}) are determined
by taking the phase fluctuations to all orders.  The density
fluctuations give no time dependence at long times, their effect
appears only in the difference between the quasi-condensate density
$\rho_{QC}$ and the total density $\rho$.

Since $D^{fb}_{\phi\phi} $ corresponds to the luminescence
spectrum, its relation to Keldysh and retarded Green's functions is as
in Eq.~(\ref{eq:56}).  Assuming that the condensation arises due to
pumping, then as in Sec.~\ref{sec:results-polar-model}, the frequency
dependence near the effective chemical potential arises from the
behaviour of the inverse retarded Green's function --- the frequency
dependence of the inverse Keldysh Green's function has no particular
singularities near this point.  In this case (which is also what is
found from the full calculations of the microscopic theory),
the singular behaviour of the $D^{fb}_{\phi\phi}$ is given by
$D^{fb}_{\phi\phi} \sim | D^R_{\phi\phi}|^2$, and so:
\begin{equation}
  \label{eq:70}
  f(t) =
  \int \frac{d \omega}{2\pi} \int \frac{k dk}{2\pi}
  \frac{(C^2/\rho) \left[ 1 - e^{-i\omega t} \right]}{
    \left| \omega^2 + 2 i \omega x - c^2 k^2 \right|^2}.
\end{equation}
(The factor of $1/\rho$ occurs from the relation of phase-phase
Green's functions to $\psi,\psi^\dagger$ Green's functions).  As one
expects for a two-dimensional system, after integrating over $\omega$,
the above integral reduces to an expression $\sim \int dk/k$, and so
one has logarithmic behaviour, cut off at high $k$ by a maximum energy
of excited modes, and at small $k$ by the time dependence.  At small
$k$, the poles of the $\omega$ integral are at $\omega=\pm 2 ix, \pm i
(c k)^2/2 x$; the first of these has a finite residue as $k \to 0$,
while the latter has a residue that is is diverging, and thus
dominates the behaviour.  The asymptotic behaviour is thus given by:
\begin{equation}
  \label{eq:71}
  f(t) =
  \int \frac{k dk}{2\pi}
  \frac{C^\prime}{ 4 x (ck)^2}
  \left[ 1 - e^{-c^2 k^2 t/2x} \right]
\end{equation}
(where $C^\prime$ is a new constant).  This expression has a cutoff
for small $k$ given by $k \sim \sqrt{x/t}/c$; thus one still has power
law correlations as for an equilibrium two dimensional gas, but with a
different power law, now set not only by the condensate density
  but also by the pumping and decay strength.  Were one to calculate
also the long-distance correlations at equal times, one would note
another difference from equilibrium.  In equilibrium, the decay of
long-time equal-position correlations, and long-distance equal-time
correlations have the same power laws.  For the spectrum in
Eq.~(\ref{eq:67}), the power-law for long-distance equal-time
correlations is twice that of long-time equal-position decay.  This is
because the low momentum cutoff for long distances is always $k \sim
1/r$, whereas the long-time cutoff is $k \sim 1/c t$ in equilibrium,
but $k \sim \sqrt{x/t}$ here.

\subsection{Finite-size effects -- lineshape of trapped system }
\label{sec:cond-linesh-finite}

\begin{figure}[htpb]
  \centering
  \includegraphics[width=3.5in]{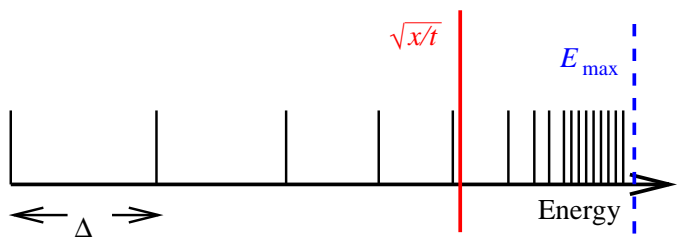}
  \caption[Energy levels in a finite system]{Spacing of discrete
    energy levels, and upper/lower cutoff energies}
  \label{fig:energy-levels}
\end{figure}

For a confined system, the integral over $k$ modes is replaced by a sum
over a discrete set of modes; i.e.:
\begin{equation}
  \label{eq:72}
  f(t)
  =
  \sum_n
  \int \frac{d \omega}{2\pi} 
  \frac{C^\prime\left[ 1 -  e^{-i\omega t} \right]}{%
    |\omega^2 + 2 i \omega x - \xi_n^2|^2}
  \simeq
  \sum_n
  \frac{C^\prime}{ 4 x \xi_n^2}
  \left[ 1 - e^{-\xi_n^2 t/2x} \right].
\end{equation}
In this form, one may then consider how the value of the sum depends
on the relative size of the mode spacing $\Delta E$, the low energy
cutoff $\sqrt{x/t}$, and the maximum energy $E_{\mathrm{max}}$.  Let
us assume the maximum energy is large, then we have a picture
something like Fig.~\ref{fig:energy-levels}.

The sum can be divided into parts above and below the low energy cutoff, giving:
\begin{equation}
  \label{eq:74}
  f(t) \simeq
  C^\prime \left[
    \sum_{n=0}^{\xi_n<\sqrt{x/t}}
    \frac{ t}{ 8 x^2 }
    +
    \sum_{\xi_n>\sqrt{x/t}}^{\xi_n=E_{\mathrm{max}}}
    \frac{1}{ 4 x \xi_n^2}
  \right]
\end{equation}
If both of these sums have many terms, then they may be approximated
by integrals, and if the density of states $\nu(\xi)$ is $\nu_0 \xi$
as it would be for a two-dimensional system with $\xi_n = c p_n$ (as
illustrated in Fig.~\ref{fig:energy-levels}) then this becomes:
\begin{eqnarray}
  \label{eq:75}
  f(t) &\simeq&
  C^\prime \nu_0\left[
    \int\limits_{0}^{\sqrt{x/t}}
    \frac{t}{8 x^2} \xi d\xi
    +
    \int\limits_{\sqrt{x/t}}^{E_{\mathrm{max}}}
    \frac{\xi d\xi}{ 4 x \xi^2}
  \right]
  \simeq
   \frac{C^\prime \nu_0}{16x} + \frac{C^\prime \nu_0}{4x}\ln\left( E_{\mathrm{max}}
      \sqrt{\frac{t}{x}} \right).
\end{eqnarray}
What is to be noted here is that the number of terms in the first part
compensates the $t$ dependence, leading to a harmless constant.  If
however the number of terms in the first term is small, or is in fact
truncated at its minimum value of one (which will inevitably occur
for large enough $t$), then one instead has something of the form:
\begin{equation}
  \label{eq:76}
  f(t) \simeq
  C^\prime  \left[ \frac{t}{8x^2} + 
    \frac{\nu_0}{4x}\ln\left( \frac{E_{\mathrm{max}}}{\Delta E} \right) \right],
\end{equation}
and so one has exponential decay of coherence at long times, arising
from the restricted number of modes.

If the mode spacing is such that only a single mode is involved, then
Eq.~(\ref{eq:72}) reduces to a single term in the sum, $\xi_0=0$, and
the result becomes very similar to the form found from models of phase
noise for a single mode condensate\cite{whittaker09,kubo54}, for which
the lineshape interpolates between Gaussian and Lorentzian:
\begin{equation}
  \label{eq:77}
  f(t) = C^\prime \int \frac{d\omega}{2\pi}
  \frac{1-e^{-i\omega t}}{(\omega^2+0^2)(\omega^2+4 x^2)}
  = \frac{C^\prime}{16x^2}\left[ 2xt - 1 + e^{-2xt}\right],
\end{equation}
hence the decay of coherence varies from $t^2$ at short time (giving
Gaussian lineshape at high frequencies) to $t$ at long times (giving a
Lorentzian peak at low frequencies).  This result is exactly as one
expects for phase noise from varying densities\cite{kubo54}:
\begin{equation}
  \label{eq:80}
  \partial_t \phi = - i U N, \quad \partial_t N = - \Gamma N + F_\Gamma(t),
\end{equation}
where $F_\Gamma$ is a Gaussian delta correlated noise noise noise with
strength $P_\Gamma$.  Solving these equations in Fourier space, one
has:
\begin{equation}
  \label{eq:81}
  \langle|\phi_\omega|^2\rangle = 
  \frac{U^2}{\omega^2} \langle|N_\omega|^2\rangle
  =
  \frac{U^2 P_\Gamma}{\omega^2(\omega^2+\Gamma^2)},
\end{equation}
which is the same form as in Eq.~(\ref{eq:77})

This section thus shows another distinction between condensates and
lasers in terms of many-mode or single mode fluctuations. If the
system is large the spatial fluctuations resulting from the continuum
of modes give rise to (in two dimensions) a power-law decay of
correlations as for an infinite, equilibrium, two-dimensional
quasi-condensate.  For smaller systems, or at longer times, the power
law crosses over to exponential decay, given by a fluctuations within
the single lowest energy mode (the other modes are too high in
energy to be relevant), as is characteristic for lasers.

\section{Summary}

This chapter has discussed in detail how the
non-equilibrium Green's function formalism can be applied to study a
model of microcavity polaritons, driven out of equilibrium by coupling
to two baths.  This model system, while not incorporating all features
of the real system, allows one to make particularly transparent
connections between laser theory and equilibrium descriptions, as well
as allowing clear illustrations of the consequences of the
approximations typically used for simple lasers.  By considering
steady states of the system in which there is a coherent photon field,
one finds a criterion for condensation to occur, and can find a
self-consistency condition which determines how the amplitude and
frequency (effective chemical potential) of the coherent field depend
on the strength of the pumping and decay.  By considering fluctuations
about steady states, one can determine whether a given steady state is
stable, find the spectrum of possible excitations, and find how this
spectrum is populated.

Starting from the normal state, without a condensate, and increasing
pumping strength, one finds that fluctuations about the normal state become
unstable at the same point that a condensed solution appears.  The
scenario by which this instability occurs on increasing pumping
strength is quite instructive.  As pumping strength increases, a
region of energies for which there is gain appears in the spectrum.
The energy dividing this region of gain from regions of loss defines
an effective chemical potential, at which the non-equilibrium
distribution function diverges.  Instability occurs at a higher
pumping strength, when this effective chemical potential (and thus the
region of net gain) reach the normal modes of the strongly coupled
system, at which point polariton condensation occurs.  Such a
description unites the lasing picture of gain exceeding loss with the
equilibrium picture of the chemical potential reaching the bottom of
the band.

While the above description allows polariton condensation to be
discussed in the language of laser theory, the results are rather
different from the normal limits assumed for a simple laser theory.
However, simpler laser theory results can be recovered within the
model discussed here, as corresponding to a high temperature limit.
In this high temperature limit, pumping corresponds to effectively
white noise, and this was shown to mean that gain only exists when
pumping bath is inverted.  This has the consequence that in this high
temperature limit, lasing and strong coupling do not coexist, whereas
they can in the low temperature polariton condensate.

When considering fluctuations about the condensed state, a somewhat
different distinction between simple lasers and the polariton
condensate emerges: the effect of finite system size, and the spectrum
of collective phase modes.  For an infinite two dimensional system,
the decay of coherence at long distances and long times is power law,
as in equilibrium (but with different powers).  For a finite system,
the effects of finite lifetime and finite size combine to lead to
exponential decay at long times.  In the limits of very small system
size, the standard result for phase noise in a single mode condensate
is naturally recovered.

To summarise, the approach presented here provides a way to connect a number
of different approaches to equilibrium and non-equilibrium condensates, 
as well as theories of lasers, in a transparent manner, allowing one to
understand the significance of various approximations,  as well as the
relations between some of the other approaches one may use.

\bibliography{keldysh-arxiv}

\end{document}